\documentclass[prd,twocolumn,showpacs,floatfix,nofootinbib]{revtex4}
\usepackage{bm}
\usepackage{amssymb}
\usepackage{graphicx}
\usepackage{epstopdf}

\begin{document}
\title{Late-time Kerr tails revisited}
\author{Lior M.~Burko$^{1,2}$ and Gaurav Khanna$^3$}
\affiliation{$^1$ Department of Physics, University of Alabama in Huntsville, Huntsville, Alabama 35899, USA\\
$^2$ Center for Space Plasma and Aeronomic Research, University of Alabama in Huntsville, Huntsville, Alabama 35899, USA\\
$^3$ Department of Physics, University of Massachusetts -- Dartmouth, N.~Dartmouth, Massachusetts 02747, USA}
\date{Version of March 19. 2008}
\begin{abstract}
The decay rate of late time tails in the Kerr spacetime have been the cause of numerous conflicting results, both analytical and numerical. In particular, there is much disagreement on whether the decay rate of an initially pure multipole moment ${\ell}$ is according to $t^{-(2{\bar\ell}+3)}$, where ${\bar\ell}$ is the least multipole moment whose excitation is not disallowed, or whether the decay rate is according to $t^{-n}$, where $n=n({\ell})$. We do careful 2+1D numerical simulations, and explain the various results. In particular, we show that  pure multipole outgoing initial data in either Boyer--Lindquist on ingoing Kerr coordinates on the corresponding slices lead to the same late time tail behavior. We also show that similar initial data specified in terms of the  Poisson spherical coordinates lead to the simpler $t^{-(2{\bar\ell}+3)}$ late time tail. We generalize the rule $n=n({\ell})$ to subdominant modes, and also study the behavior of non--axisymmetric initial data. We discuss some of the causes for possible errors in 2+1D simulations, demonstrate that our simulations are free of those errors, and argue that some conflicting past results may be attributed to them.
\end{abstract}
\maketitle

\section{Introduction and summary}

The late-time tails of black holes have been studied in much detail since Price's seminal work \cite{price}. The formulation of the problem is a straightforward one: place an observer in a circular orbit around a black hole, and have her measure at late times a generic perturbation field, that had compact support at some initial time. It is generally accepted that the observer measures the late-time perturbation field to drop off as an inverse power law of time, specifically as  $t^{-n}$. It is the value of $n$ that has been controversial in the literature, with some conflicting results reported. 

In the case of a Schwarzschild black hole, $n=2\ell+3$, where $\ell$ is the multipole moment of the initial perturbation field. Namely, if the initial (compactly supported) perturbation field has the angular dependence of $Y_{\ell}^m$, the angular dependence remains unchanged (``spherical harmonics are eigenvectors of the Laplacian operator"), and the decay rate of the field is governed by the $\ell$ value of the initial perturbation. These results remain unchanged also for a Reissner--Nordstr\"{o}m black hole \cite{bicak} (including the extremal case \cite{blaksley}), that like the Schwarzschild black hole is spherically symmetric. Notably, a generic perturbation is a linear superposition of $\ell$-modes, so that the tail is dominated by the slowest damped mode, or by the lowest value of $\ell$.

For rotating black holes conflicting values of $n$ have been reported. Obviously, the observer measures a well defined decay rate. Of the various conflicting values in the literature some are wrong (i.e., they report a value inconsistent with the outcome of the initial value problem), while other results are correct (in the same sense, of reporting the right value for the initial value problem). As it turns out, different initial value formulations may lead to different decay rate for the tails. While none of these is ``more correct" {\em sensu stricto} than others, it is important to understand the relationship of different initial value sets, so that the understanding may lead to some insights. 

One of the correct answers to the question of the value of $n$ was described, e.g., in \cite{scheel}: the late-time field decays according to the same rule as in the spherically-symmetric case, namely according to $t^{-(2{\bar\ell}+3)}$, where ${\bar\ell}=m$ if $\ell-m$ is even, and ${\bar\ell}=m+1$ if $\ell-m$ is odd, independently of the initial value of $\ell$. Below, we dub this tail as ``Tail A." The reason for this decay rate is that spherical harmonic modes do not evolve independently on the background of a rotating black hole (``spherical harmonics are {\em not} eigenvectors of the Laplacian operator"). Starting with an initial value for $\ell$, infinitely many other $\ell$ modes are excited, respecting only the dynamical constraints ($\ell\ge s$, where $s$ is the spin of the field, and $\ell\ge |m|$) and the equatorial symmetry of the initial data (such that even and odd modes do not mix). 

Another correct answer to the question was discussed recently in \cite{GPP} and in \cite{TKT}. According to this result, 
$n=2\ell +3$ if $\ell -m<2$, $n=\ell +m+1$ if $\ell -m\ge 2$ is even, and $n=\ell+m+2$ if $\ell -m\ge 2$ is odd. This result is dubbed below as ``Tail B." Evidently, this result is inconsistent with that of Tail A. The Tail B behavior was first predicted in \cite{hod}. While either the Tail A or the Tail B behavior is a solution to {\em some} initial value problem, these problems are different enough to lead to markedly different late--time evolutions. 

In this paper we present some explanation for these different results, that resolves the apparent discrepancy at least for some cases. This paper's motivation is to explain some of the historical discrepancies, in what we believe is an illuminating manner. Specifically, finding the reasons for the controversy in the literature teaches us important lessons. In particular, we consider two results, conflicting both with each other and with the results of this paper: the numerical results of \cite{krivan}, and the analytical model of \cite{poisson}. More precisely, we consider the simplest case for which there is a disagreement in the literature, specifically the case of a massless scalar field ($s=0$) perturbation of a Kerr black hole, with initial $\ell=4$ and $m=0$. For this case, the late time tail drops as $t^{-3}$ according to Tail A, because the monopole $(\ell=0$) mode of the scalar field can and is excited, and this mode's drop off rate obeys the $t^{-(2\ell+3)}$ rule. According to Tail B, the late--time decay rate is $t^{-5}$. The (numerical) results of \cite{krivan} suggest a drop off rate of $t^{-5.5}$ that does not appear to converge to the decay rate of Tail B. The Poisson analytical model (of a globally weakly curved spacetime) predicts a drop off rate of $t^{-5}$ if carried over to the Kerr case \cite{poisson}, which may suggest a Tail B behavior. However, as we show below, the correct behavior of such tails should be that of Tail A. 

We discuss in detail the tails developing from initial data specified on ingoing Kerr silces of \cite{burko-khanna}. It was suggested in \cite{GPP} that the Tail A results of \cite{burko-khanna} are due to the slicing condition. We show that this is not the case. In fact, one can find initial condition sets with the same slicing condition as in \cite{burko-khanna} (slicing corresponding to the ingoing Kerr coordinates) that result in either Tail A or Tail B decay rates, in just the same way as with Boyer--Lindquist slicing.  In fact, although the Boyer--Lindquist  and ingoing Kerr time slices do not coincide, we find that similar initial data lead to the same tail behavior in either. We therefore suggest that different slicing conditions may fall into equivalency classes, and Boyer--Lindquist and ingoing Kerr slices belong to the same class. We propose that this is the case because of the properties of the transformation from Boyer--Lindquist and ingoing Kerr coordinates. 

In the case of \cite{krivan}, we attribute the (numerically stable!) result of a drop off rate of $t^{-5.5}$ to an intermediate tail being misidentified as an asymptotic tail. As we show, the traditional representation of the tail in a log--log plot of the field as a function of time is susceptible to lead to an ``optical illusion," that may support this misidentification. Specifically, taking initial conditions that lead to a Tail A decay rate, we show what may be interpreted as a tail falling off as $t^{-5.5}$, similar to what is found in \cite{krivan}, except that if the integration is carried on further in time the intermediate behavior is taken over by the true asymptotic behavior, or a tail falling off as $t^{-3}$. A similar effect can also be found for the Tail B case. The reason \cite{krivan} mistook an intermediate tail for an asymptotic one is that \cite{krivan} chose parameters that make it harder to distinguish the two using the crude tools used in \cite{krivan}. Specifically, we show that the farther out the initial pulse, the longer the intermediate tail. Reference \cite{krivan} chose the initial pulse very far out, which resulted in a protracted intermediate tail. While this choice made the problem severe enough, it was even further exacerbated by making the initial pulse outgoing and by taking the Kerr parameter $a/M=0.9999$. Either choice makes the beginning of the tail regime postponed. As the integration time in \cite{krivan} is rather short, we conclude that the tail analyzed in  \cite{krivan} is an intermediate one, and that if the integration time were longer, \cite{krivan} too would have found a $t^{-3}$ tail (for Tail A initial data) or a $t^{-5}$ tail (for Tail B initial data). (In view of our argument below, we comment that the code of \cite{krivan} was fourth order angularly, which we find to be satisfactory for the relatively short runs done by \cite{krivan}.) 

How can one then avoid misinterpreting intermediate tails as asymptotic ones? We provide a criterion, that we believe can be used for the identification of the asymptotic tail. By our criterion, a tail is asymptotic if the local power index \cite{burko-ori} is linear in $t^{-1}$ both locally and globally, in a sense that we make precise. 

In the case of \cite{poisson}, we explicitly follow the proposal of Poisson, and evolve initial data that are a pure spherical harmonic mode with respect to ``spherical" coordinates, not the Boyer--Lindquist coordinates, that are ``spheroidal" in the language of \cite{poisson}. According to \cite{poisson}, such initial data should lead to a Tail B behavior when applied to a Kerr background. We show that this is not the case, and Tail A behavior is found. Specifically, starting with $\ell=4$ and $m=0$, Poisson predicts the asymptotic tail to drop off as $t^{-5}$. In \cite{poisson}, such time evolution is demonstrated in a globally weak--curvature spacetime (that has no symmetries), and it is conjectured that such behavior is carried over to the case of a Kerr black hole. We show that this conjecture is not realized: instead, the evolution is according to Tail A, and the drop off rate is according to $t^{-3}$. 

Very recently, Gleiser, Price, and Pullin (henceforth GPP) have proposed a theoretical model that expands the wave equation in (even) powers of the black hole's spin rate $a$ \cite{GPP}. The advantage of the GPP approach is that the zeroth order wave operator is identical to the spherically--symmetric operator, so that it can be solved numerically using a simple 1+1D code. Each iteration in the GPP approach gives the field to the next order in $a^2$, using the field of the previous iteration as the source term of a spherically symmetric wave operator. GPP argue that their approach is superior to the 2+1D approach, because it frees the numerics from the subtleties of the 2+1D simulations, and because the 1+1D numerics are inherently more reliable. We  agree with these claims. 
The GPP argument implies that delicate cancellations exactly cancel not only the leading tail term (in an expansion in $t^{-1}$), but also the leading subdominant mode \cite{smith-burko} (for an initial $\ell=4$; for other initial choices more subdominant modes may need to be cancelled). GPP further argue that turning off the coupling term at very late times destroys the exact cancellation, so that the tail returns to the ``naive" decay rate. It is quite remarkable that even if this is done at extremely late times, so that the hexadecapole field is much smaller than the monopole field, it still does not lose its ability to exactly cancel out its two leading terms (in $t^{-1}$).  

Notably, GPP imply that different slicing conditions may be responsible for the apparently conflicting results of numerical simulations. Indeed, in \cite{burko-khanna} the slicing condition was that associated with ingoing Kerr coordinates, in \cite{scheel} the slicing was that of the Kerr--Schild coordinates, and in \cite{krivan} Boyer--Linquist slicing was used. GPP argue that the translation of an initially pure $\ell=4$ mode, say, in some other slicing condition (e.g., ingoing Kerr or Kerr--Schild)  into Boyer--Lindquist slicing results in a mixed state of modes for the Boyer--Lindquist slicing, and the existing lower $\ell$ (specifically, the quadrupole and monopole) modes on the initial time slice are the ones responsible for the tail structure. According to GPP, among all slicing conditions the Boyer--Lindquist one is special, as only in this slicing an initial pure $\ell=4$ mode leads to a tail of $t^{-5}$, and in all other slicing conditions an initially pure $\ell=4$ mode leads to a $t^{-3}$ tail. We disagree with this viewpoint, as we find that also for ingoing Kerr slicing the tail decays like $t^{-5}$. 

More recently, Tiglio, Kidder, and Teukolsky (TKT) \cite{TKT} performed careful 2+1D simulations of the axisymmetric Kerr tail problem using a pseudospectral code. The pseudospectral approach has an advantage over the finite differencing approach, in that the convergence is exponential in the former, and only a power law in the latter. Consequently, in the pseudospectral approach the noise evolution problem that permeates finite differencing computations is nearly nonexistent. TKT studied two slicing cases: Kerr--Schild slicing and Boyer--Lindquist slicing. TKK found the Tail A behavior in the former (consistently with the conclusions of \cite{scheel} that were obtained in 3+1D) and the Tail B behavior in the latter (consistently with GPP). TKT correctly point out the main reason for the difference between Tail A and Tail B behaviors that are found in Boyer--Lindquist initial data and on Boyer--Lindquist slicing, specifically that the zero ``field momentum" of the former is in fact a mixed state of pure multipole modes. We consider this question in great detail, and show that the momentum condition is a very special combination of states, as described below. 

We discuss in detail the noise evolution problem that may cause a Tail B masquerade as a Tail A. That is, numerical noise evolves on the grid the same way that the physical signal does, so that the Tail A behavior of the noise evolution overwhelms the Tail B behavior of the physical field. We then show that with care this phenomenon can be easily identified. 

We  next discuss how initial data with the same slicing condition could lead to either Tail A or Tail B behavior. As shown first by GPP and then by TKT, initial data that are a pure multipole of the angular Boyer--Lindquist coordinates in Boyer--Lindquist slicing lead to Tail B behavior. When one sets the initial conditions by requiring that the field's ``momentum" $\Pi$ (see below for definitions) vanishes, the tail's behavior is that of Tail A. Indeed, the condition on the momentum implies that the field's multipole structure is changed, as the field's velocity now is proportional to the field multiplied by a function of the angular coordinate $\theta$. Restricting attention to low black hole spin values ($a/M\ll 1$), this function is proportional to 
$a^2\,\sin^2\theta$. Therefore, one would expect that in addition to an initial hexadecupole at $O(a^0)$ one would have now also all other (even) multipoles at $O(a^2)$. An existing monopole on the initial slice at $O(a^2)$ would then lead to a tail that has the same decay rate as a pure multipole that behaves like Tail A, which is at $O(a^2)$. However, we show that the tail is at $O(a^4)$, not $O(a^2)$. The explanation is that the leading contribution (in $a^2$) does not contribute to the monopole, so that on the initial slice the monopole is at $O(a^4)$, not at $O(a^2)$. Evolution of this mode then can dominate at late times, showing a tail that is both at $O(a^4)$ and has the Tail A behavior. Nevertheless, this is not the only contribution to this tail. The initial data contains also a quadrupole at $O(a^2)$, which also evolves a $O(a^4)$ monopole because of the mode excitation effect. However, as in this case the quadrupole has an $\ell$ value that differs from the monopole's by only $2$, the behavior is again consistent with that of Tail A. Therefore, the Tail A behavior has three sources: a pre-existing monopole at $O(a^4)$, a pre-existing quadrupole at $O(a^2)$, and a pre-existing hexadecapole at $O(a^0)$, all three behaving consistently with Tail A. 

We further extend the numerical study to non-axisymmetric modes, and find that the Tail A and Tail B behaviors occur for such modes consistently with their behavior in the axisymmetric case, i.e., they obey the aforementioned decay rate rules. We then extend the argument for Tail B behavior to subdominant modes. Specifically, we argue that starting with an initial mode $\ell$, the decay rate of an (allowed) excited mode $\ell'$ is given by $n=\ell+\ell'+1$ if $\ell-\ell'>2$, and $n=2\ell'+3$ otherwise. This rule includes that of \cite{hod} as a particular case, when $\ell'=m,m+1$. In particular, we consider the axisymmetric case of  $\ell=6$ and $\ell'=2$ (which is a subdominant mode, as also the monopole is excited). The quadrupole mode $\ell'=2$ then indeed decays with $n=9$ as predicted by our proposal.

We are hopeful that this Paper may put some order in a confused state of affairs, in which the literature includes side by side correct results and incorrect ones, without a proper explanation of the incorrect ones, and also apparently conflicting correct results without a proper explanation of the sources for the apparent conflict. 

The organization of this paper is as follows. In Section \ref{reliability} we discuss the reliability of the 2+1D approach in calculating tails, and how carelessness can result in an incorrect value for the field. In Section \ref{a2} we discuss the decomposition of the monopole projection of the tail in powers of $a^2$, and in Section \ref{optical_illusion} we discuss one possible pitfall that can cause a correct field be misinterpreted. In Section \ref{s_ingoing_kerr} we study tails in ingoing Kerr coordinates, and in Section \ref{poisson_data} we analyze a pure mode in ``spherical" coordinates \cite{poisson}, and show that it leads to the same tail decay rate as in ``spheroidal" coordinates. 

\section{Reliability of the 2+1D code}\label{reliability}

The solution of the Teukolsky equation in 2+1D is delicate and sensitive, and can lead to qualitatively wrong results if proper care is not taken. In Section \ref{optical_illusion} below we show that even when the numerically computed field is the correct one, careless interpretations may lead to incorrect conclusions. In this section we first demonstrate how carelessness may lead to qualitatively wrong behavior of the solution, and then we demonstrate the reliability of the approach when proper care is taken. 

\subsection{Second order angular differentiation}

The standard of scientific computation in the finite difference approach has been second order algorithms for many years. Indeed, second order convergence typically implies the solution is both consistent and stable to a practical level, so that the solution is accurate enough without having to reduce the typical grid spacing to extremely low levels that would make the code too costly computationally. Normal application of the Dahlquist theory (extended to PDEs) then means that if the discrete formula approximating the differential equation has locally a positive order of accuracy (convergence), and that if numerical errors do not grow unboundedly (stability), then the solution is convergent, in the sense that as the grid spacing goes to zero, the solution would approach the correct one. As we show below, the tail problem involves the Dahlquist theory with a twist. 

Early codes to calculate the late--time Kerr tail were therefore designed to be second--order in all coordinates, i.e., temporally, radially, and with respect to the angular coordinate $\theta$ \cite{KLPA}. (In the 2+1D approach the field is decomposed into $\varphi$ modes, so that one does not have numerical derivatives with respect to $\varphi$.) While this approach seems very natural, it can mislead into qualitatively wrong solutions. Consider first a Schwarzschild black hole, excited by an azimuthally  symmetric ($m=0$) field with $\ell=4$. The decay rate of the Price tail is incontrovertible in this case: as the background is spherically symmetric, spherical harmonic modes evolve independently, and the decay rate is therefore $t^{-11}$. Indeed, 1+1D codes readily verify this result. 

\begin{figure}[htbp]
   \includegraphics[width=3.4in]{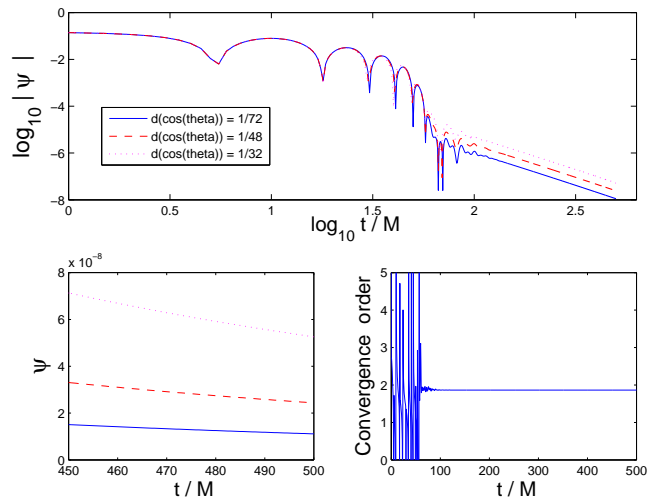} 
\caption{Convergence tests for Schwarzschild initial data with the second--order code. Upper panel and lower panel on the left: the full field (evaluated at $\theta=\pi/2$) for three different angular resolutions ($\,\Delta\cos\theta=1/72$ (solid), $1/48$ (dashed), and $1/32$ (dotted)) as a function of time. The lower panel on the right is the convergence order as a function of the time at $r_*/M=25$. 
The initial data are for a  Schwarzschild black hole ($a/M=0$), and are a gaussian centered at $r_*=25M$ with width $4M$. The radial resolution is $\,\Delta r=M/32$, and the temporal resolution is chosen to be $\,\Delta t=\,\Delta r/2$, which satisfies the Courant condition. The initial data are a pure multipole with 
$\ell=4$.}
\label{2nd_1}
\end{figure}

The numerical solution with a second-order 2+1D numerical code may lead however to the erroneous conclusion that the decay rate is $t^{-3}$: in Fig.~\ref{2nd_1} we show the late--time field as a function of time for three choices of the angular grid spacing, and for fixed radial and temporal grid spacings. As can be readily inferred from the upper panel in Fig.~\ref{2nd_1}, the decay rate is $t^{-3}$, which is definitely wrong. The origin of this wrong result is indicated in the same figure: the field values themselves reduce considerably in amplitude under refinement of the grid (almost as the second power of the ratio of the grid parameters). The signal seen is therefore not the physical field, but an evolved numerical noise (``leakage of multipoles"). Our interpretation is that while the physical signal indeed drops like $t^{-11}$, it is swamped by the evolved numerical noise, which has a monopole component. The latter drops like $t^{-3}$, so that {\rm any} very long time evolution must be dominated by numerical noise. Refinement of the grid parameter can postpone the numerical-noise dominated regime, but cannot utterly eliminate it: refinement of the grid parameter reduces the level of the numerical noise, so that if the grid is fine enough, the noise level at a particular value of the time reduces to below the level of the physical field. However, as the former drops more slowly than the latter, at very late times the field is dominated by numerical noise (see Fig.~\ref{6th_1} below). 

Here enters the twist on the Dahlquist theory: the numerical noise does not grow unboundedly. In fact, it even drops quadratically with the grid spacing, and drops with time. It does so, notwithstanding, slower than the physical signal, so that the signal may be dominated by numerical noise even though the regular Dahlquist criteria are apparently satisfied.

\subsection{Sixth order angular differentiation}

While a second order code may in principle be sufficient to see physical tails (for very fine grids), an alternative approach is to increase the order of the angular differentiation. We first show how the same initial data and grid resolutions lead to qualitatively different results in Schwarzschild, and then we discuss Kerr. 

\begin{figure}[htbp]
 \includegraphics[width=3.4in]{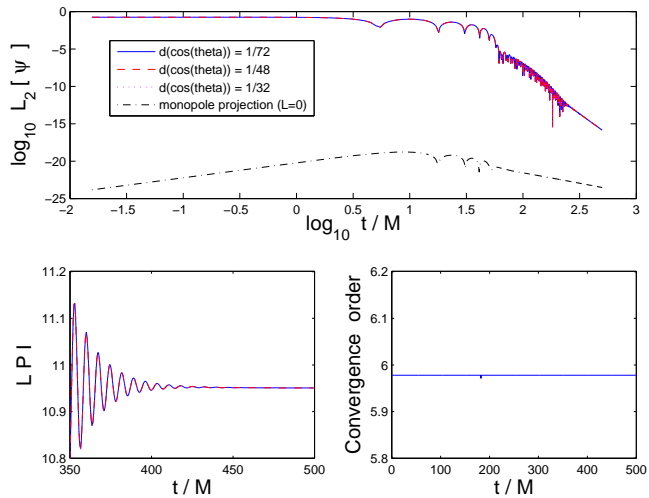} 
\caption{Same initial data and resolutions as in Fig.~\ref{2nd_1} for the sixth order code. The upper panel shows the $L_2$ norm for the full field (integrated over a sphere at $r_*=25M$) in three resolutions. The three curves are nearly indistinguishable on this plot's scale. The upper plot also shows the monopole projection of the field, which is pure evolved numerical noise for resolution of $\,\Delta\cos\theta=1/64$. This figure also shows the tails local power index as a function of time (lower left) and the convergence order of the three resolution shown on the upper panel (lower right).}
\label{6th_1}
\end{figure}

Figure \ref{6th_1} shows the sixth order convergence of the field, for the same parameters that were shown in Fig.~\ref{2nd_1} with the second order code. The field is not made of evolved noise (``leakage of multipoles"), as it clearly converges to a non-zero value under grid refinement. Notably, the field's amplitude even {\em increases} slightly under grid refinement. The hallmark of noise signals is that their amplitude {\em decreases} under grid refinement, as is indeed evident in Fig.~\ref{2nd_1}. Most importantly, the Price tail drops as $t^{-11}$, as it should. We emphasize that these conclusions are true up to some maximal time value. For longer times the residual monopole noise, that drops slower than the physical field, will inevitably dominate. Simple extrapolations reveal that the signal noise will dominate over the physical signal for $t\gtrsim 4,380M$ for our choice of angular resolution. Longer evolutions than that will be dominated by noise evolution. 
As noted above, it is not difficult to separate the physical signal regime from the noise signal regime by doing global convergence tests. It is therefore crucial {\em for any tail simulation} that convergence tests are done throughout the entire computational domain, and not limited just to short integration times (despite the temptation to save computer time --- e.g., see \cite{krivan}).   

\begin{figure}[htbp]
   \includegraphics[width=3.4in]{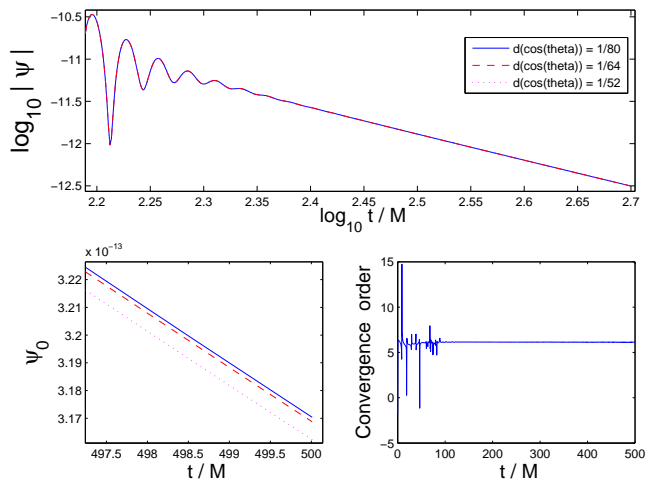} 
\caption{Same as Fig.~\ref{2nd_1} for the monopole projection of the full field using the sixth order code for Kerr. The angular resolutions are $\,\Delta\cos\theta=1/80$ (solid), $1/64$ (dashed), and $1/52$ (dotted). The initial data are an outgoing Gaussian pulse located at $r_*=25M$ with width of $4M$ with compact support on the Boyer--Lindquist  $t=0$ hypersurface of a Kerr black hole with $a/M=0.995$. The initial multipole is a pure Boyer--Lindquist $\ell=4$.  The radial resolution is $\,\Delta r=M/50$, and the temporal resolution is chosen so that $\,\Delta t=\,\Delta r/2$.}
\label{6th_2}
\end{figure}

In Fig.~\ref{6th_2} we demonstrate the sixth order angular convergence of Kerr evolutions. This figure summarizes the main parts for our claim that our tail results are reliable: The code is convergent to a non-zero value, and even increases in magnitude with grid refinement (so that the signal seen is not an evolved noise signal), and the convergence order is global, throughout the computational domain. Notice that in Fig.~\ref{6th_2} we study the convergence with the monopole projection of the total field. 
This projection is more sensitive to the created and evolved monopole numerical noise than the full field, say, as we can test its convergence also at times earlier than the instant the monopole dominates the total field for the first time.  Figure \ref{6th_3} depicts the ($L_2$ norm of the) full field and its monopole projection. The parameters are chosen so that intermediate tails are minimized, so that computation time is used efficiently.  
The monopole projection amplitude is many order of magnitudes higher than typical ``multipole leakage" noise levels (cf.~with Fig.~\ref{6th_1}). Indeed, it is the result of physical multipole excitations, not noise generation. 
Below, we also bring evidence that the observed tail is indeed asymptotic, so that we are not misled by an intermediate tail masquerading as an asymptotic one.

\begin{figure}[htbp]
   \includegraphics[width=3.4in]{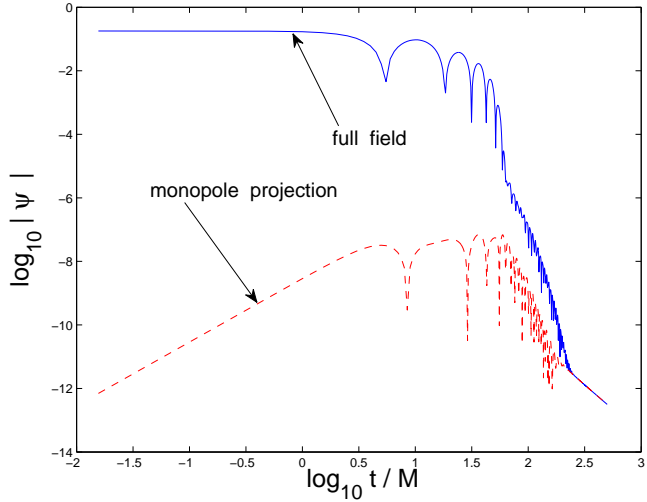} 
\caption{Same as Fig.~\ref{6th_2}. Shown are the $L_2$ norm of the full field (solid curve) and the field's monopole projection (dashed curve). For both, the angular resolution is $\,\Delta\cos\theta=1/72$ and the radial resolution is $\,\Delta r=M/32$. The temporal resolution is chosen so that $\,\Delta t=\,\Delta r/2$. }
\label{6th_3}
\end{figure}

\section{The tail in an $a^2$ expansion}\label{a2}

\begin{figure}[htbp]
   \includegraphics[width=2.8in]{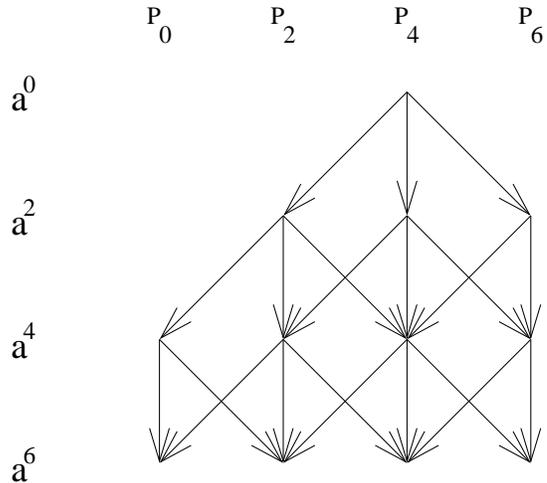} 
\caption{The mode coupling schematics: at each iteration in $a^2$ in the GPP approach, each $\ell$ modes excites three modes, specifically $\ell-2,\ell$, and $\ell+2$. The initial data are purely hexadecapolar. We show only modes up to $O(a^6)$ and $\ell=6$.}
\label{6th_2_1}
\end{figure}

GPP argue that when expanded in powers of $a^2$, the tail would be at $O(a^4)$. When a  mode $P_n$ exists at any iteration order (in $a^2$), at the next iteration three modes will result: $P_{n-2}$ (if $n-2\ge 0$), $P_n$, and $P_{n+2}$. The mode excitation scheme is shown in Fig.~\ref{6th_2_1}. In the GPP approach, this behavior comes about because of the term $\sin^2\theta\times P_n(\cos\theta)$ appearing in the source term (see \cite{GPP} for detail).  This way, starting with a pure $\ell=4$ mode, there is just one way to create a monopole field at $O(a^4)$, but three ways to create a monopole field at $O(a^6)$. 
We study two classes of initial data sets: In Class A the field $\psi$ is a pure multipole (in practice, we focus attention here mostly on the azimuthal $\ell=4$ mode) and the field's ``momentum" $\Pi=0$, where $\Pi:=\,\partial_{t}\psi+b\,\partial_{{\hat r}_*}\psi=0$.  Here, 
$b=({\hat r}^2+a^2)/{\hat\Sigma}$, ${\hat\Sigma}^2=({\hat r}^2+a^2)^2-a^2{\hat\Delta}\,\sin^2{\hat\theta}$, and ${\hat\Delta}={\hat r}^2-2M{\hat r}+a^2$ (for details see, e.g., \cite{pazos}). The ${\hat r},{\hat \theta}$ coordinates are the Boyer--Lindquist ones. 
Class A initial data lead to Tail A behavior (i.e., to $n=3$), with the tail at $O(a^4)$. In Class B initial data we choose the field $\psi$ in the same way as in Class A, but replace the condition $\Pi=0$ with 
$\,\partial_{t}\psi=\pm \,\partial_{{\hat r}_*}\psi$ or $\,\partial_{t}\psi=0$ or a linear combination thereof (with coefficients that are independent of $\hat \theta$).  
Class B initial data lead to Tail B behavior (i.e., to $n=5$), with the tail at $O(a^4)$. (The momentary stationary data case  $\,\partial_{t}\psi=0$ is unique, and leads to $n=6$ in a similar way to the Schwarzschild case \cite{Price_Burko}.) The different sets of initial data are summarized in Table \ref{table2}.

\begin{table}[htdp]
\caption{The initial data classes used in the text. Unaccented coordinates mean any of the coordinate systems described in Table \ref{table1}. See the text for more detail.}
\begin{center}
\begin{tabular}{||c|c|c||}
\hline
Initial data class  & Condition & Connection \\
\hline\hline
Class A & at $t=0$: $\psi$, $\Pi=0$ & $\Pi=\,\partial_{t}\psi+b\,\partial_{{ r}_*}\psi$ \\
\hline
Class B & at $t=0$: $\psi$, $\,\partial_t\psi$ &   \\
\hline\hline
\end{tabular}
\end{center}
\label{table2}
\end{table}%

From Fig.~\ref{6th_2_1} is it evident why the late time tail for Class B data is at $O(a^4)$: The initial hexadecapole field at $O(a^0)$ excites a quadrupole field at $O(a^2)$, which in itself excites a monopole field at $O(a^4)$. Higher powers in $a^2$ are also present in the monopole field, but for small values of $a^2$ they are negligibly small compared with the leading 
$a^4$ term. Even more so, the monopole tail appears to depend very weakly on terms at $O(a^6)$: the tail turns out to be 
a predominantly $a^4$ effect.

Why then is the tail for Class A initial data also lead to a tail at $O(a^4)$? The condition $\Pi=0$ is equivalent to the condition 
$\,\partial_{t}\psi=-b \,\partial_{{\hat r}_*}\psi$. As the function $b$ depends on ${\hat \theta}$, it changes the multipolar structure of the initial data sets. Specifically, for small $a^2$, $b=1+(r-2M)\,a^2\,\sin^2{\hat\theta} / (2{\hat r}^3)+O(a^4)$. One may therefore naively expect these initial conditions to lead to a mixed state of multipoles at $O(a^2)$, that may include a monopole term at $O(a^2)$. If such a monopole term existed, the tail would be an $O(a^2)$ effect. 

To find the order (in $a^2$) of the tail we write the monopole projection of the field $\psi_0$ as an expansion in the spin parameter $a$:
\begin{equation}
\psi_0=\sum_{n=0}^{\infty}C_n(t)\times a^n\, .
\end{equation}
Class B initial data lead to only even powers of $a$ having non-zero coefficients, and $C_0(t)=0=C_2(t)$, so that the first non-zero coefficient is $C_4(t)$. For the Tail B behavior found by GPP, $C_4(t)\sim t^{-5}$ (for a pure Boyer--Lindquist initial mode with initial $\ell=4$). 

\begin{figure}[htbp]
   \includegraphics[width=3.4in]{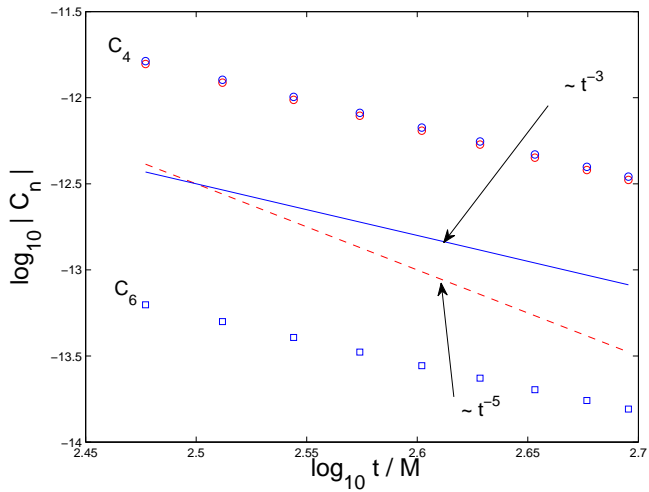} 
\caption{The coefficients $C_4(t)$ and $C_6(t)$. The two sets of data points ($\circ$) corresponding to $C_4(t)$ relate to fits to the ansatz $\psi (t) = C_4(t)\,a^4$ (slightly higher) and $\psi (t) = C_4(t)\,a^4 + C_6(t)\,a^6$.  The coefficients $C_6(t)$ are represented by $\square$. We also show two reference lines, with slopes of $-3$ (solid) and of $-5$ (dashed). 
}
\label{6th_3}
\end{figure}

For Class A initial data we can find the coefficients $C_n(t)$ from a 2+1D simulation in the following way: 
We take the monopole projection value of the field in the tail regime at a fixed value of time for various 
$a$ values. In practice, we considered $a$ values in the range $0.05\le a/M\le 0.995$. 
We then fit the data to two ansatzs: first, to  $\psi_0 = C_4 \,a^4$, and second, to $\psi_0 = C_4\,  a^4 + C_6\, a^6$. That is, we obtain the numerical values of the coefficients $C_n$ for that particular value of the time. The fit is very good: we get the squared correlation coefficient to be better than $0.999$.  We then repeat for a number of time values, so that we have $C_4(t)$ and $C_6(t)$. We can then find the time dependence of $C_4(t)$ and of $C_6(t)$. Our results are shown in Fig.~\ref{6th_3}. Both $C_4(t)$ and $C_6(t)$ drop off as $t^{-3}$. We emphasize that the raw data are taken from Fig.~\ref{6th_2}: the data converge at sixth order angularly, and is the physical signal, not an evolved noise signal. 

We therefore make the following important inferences: first, initial data of Class A lead to to Tail A behavior. Second, the tail is a $O(a^4)$ phenomenon, and $O(a^6)$ contributions are negligibly small. We checked this result not only for small values of $a^2$, but also for  values as high as $a^2\sim 0.99$. The reason for this behavior is as follows: the mixed state of multipoles at $O(a^2)$ in the initial data includes only multipoles higher than the quadrupole. The absence of the monopole is evident from the vanishing of 
$$\int_0^{\pi} P_0(\cos{\hat\theta})\times \left[\,\sin^2{\hat\theta}\,P_4(\cos{\hat\theta})\right]\,\sin{\hat\theta}\, d{\hat\theta}=0\, . $$
We do get however a quadrupole contribution at $O(a^2)$, as 
$$\int_0^{\pi} P_2(\cos{\hat\theta})\times \left[\,\sin^2{\hat\theta}\,P_4(\cos{\hat\theta})\right]\,\sin{\hat\theta}\, d{\hat\theta}=-\frac{8}{105}\ne 0\, . $$

The monopole term on the initial data appears only at $O(a^4)$. The function $b$ is 
\begin{eqnarray*}
b&=&1+\frac{{\hat r}-2M}{2{\hat r}^3}\,a^2\,\sin^2{\hat\theta} \\
&-&\frac{a^4\,\sin^2{\hat\theta}}{8{\hat r}^6}\left[ 3\,\cos^2{\hat\theta}\,({\hat r}-2M)^2+({\hat r}-2M)^2-16M^2
\right]\\
&+&O(a^6)\, ,
\end{eqnarray*}
so that the term independent of ${\hat\theta}$ inside the square brackets does not contribute to the monopole at $O(a^4)$, but the term proportional to $\cos^2{\hat\theta}$ does, because 
$$\int_0^{\pi} P_0(\cos{\hat\theta})\times \left[\,\sin^2{\hat\theta}\,\cos^2{\hat\theta}\,P_4(\cos{\hat\theta})\right]\,\sin{\hat\theta}\, d{\hat\theta}=-\frac{16}{315}\, ,$$
which does not vanish.

We therefore find that the mixed state of multipoles is such that there is a monopole at $O(a^4)$, a quadrupole at  $O(a^2)$, and a hexadecapole at  $O(a^0)$. These multipoles are arranged along a diagonal in Fig.~\ref{6th_2_1}, so that they track the mode coupling pattern, and therefore lead to a tail at $O(a^4)$, as each component of the mixed state of multipoles in itself leads to the same tail. Notably, when the initial data correspond to $\ell=6$ the modes of the mixed state are again arranged along a diagonal, with a hexindatesserapole at $O(a^0)$, a hexadecapole at $O(a^2)$, a quadrupole at $O(a^4)$ and a monopole 
at $O(a^6)$.  This behavior remains also for an initial $\ell=8$ data, with all mixed terms arranged along a diagonal in Fig.~\ref{6th_2_1}. We therefore expect it to be a general result, and propose that Class A initial data tails are at the same order in $a^2$ as Class B tails.

\section{Evolution of Class B initial data}

\subsection{Axisymmetric modes}

\begin{figure}[htbp]
   \includegraphics[width=3.4in]{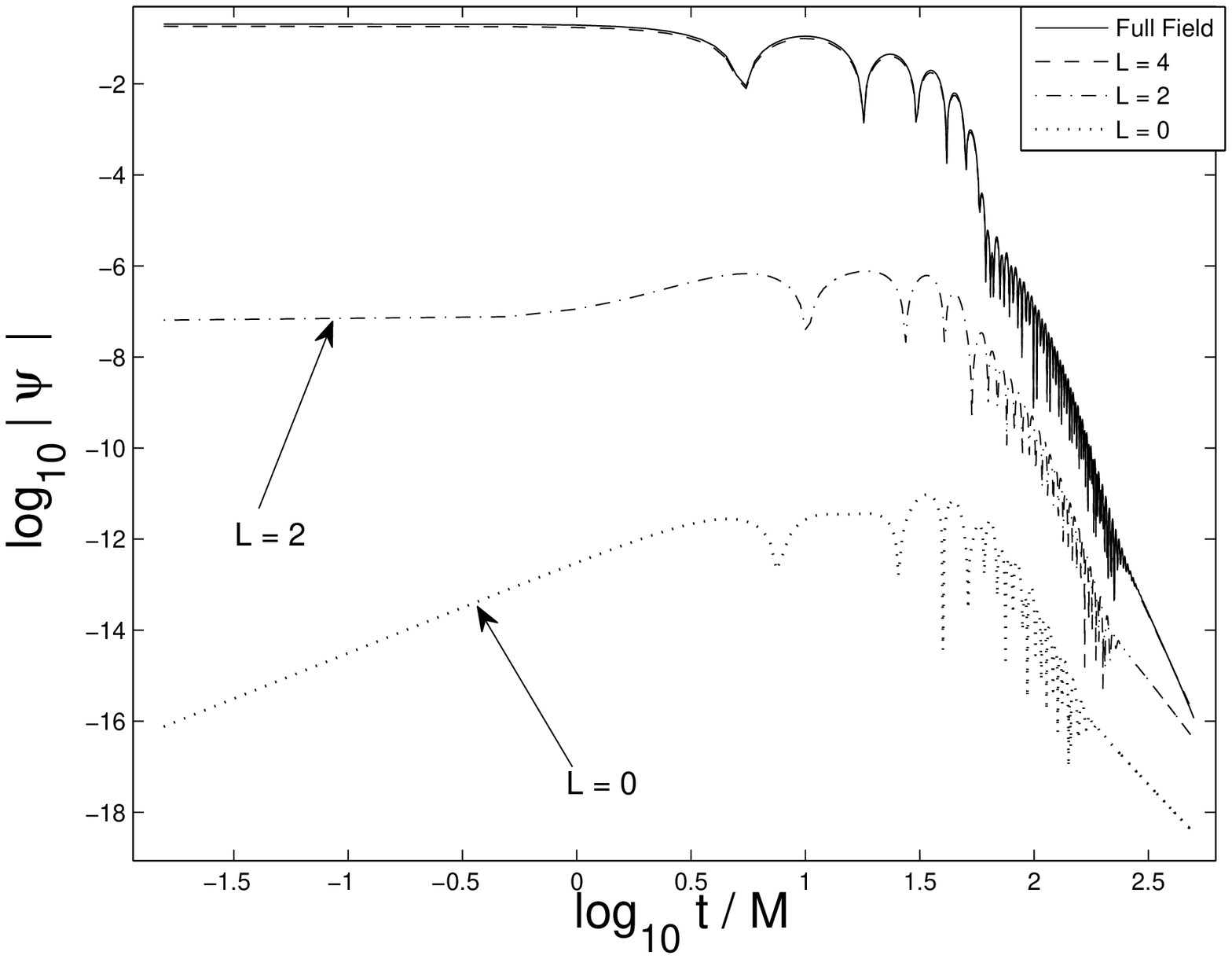} 
\caption{Same initial data as in Fig.~\ref{6th_2} but for Class B: The gaussian pulse is centered at $r_*=25M$ with width of $4M$.  Grid parameters are: $\,\Delta\theta=1/48$, $\,\Delta r_*=M/32$.}
\label{classB_4field}
 \includegraphics[width=3.4in]{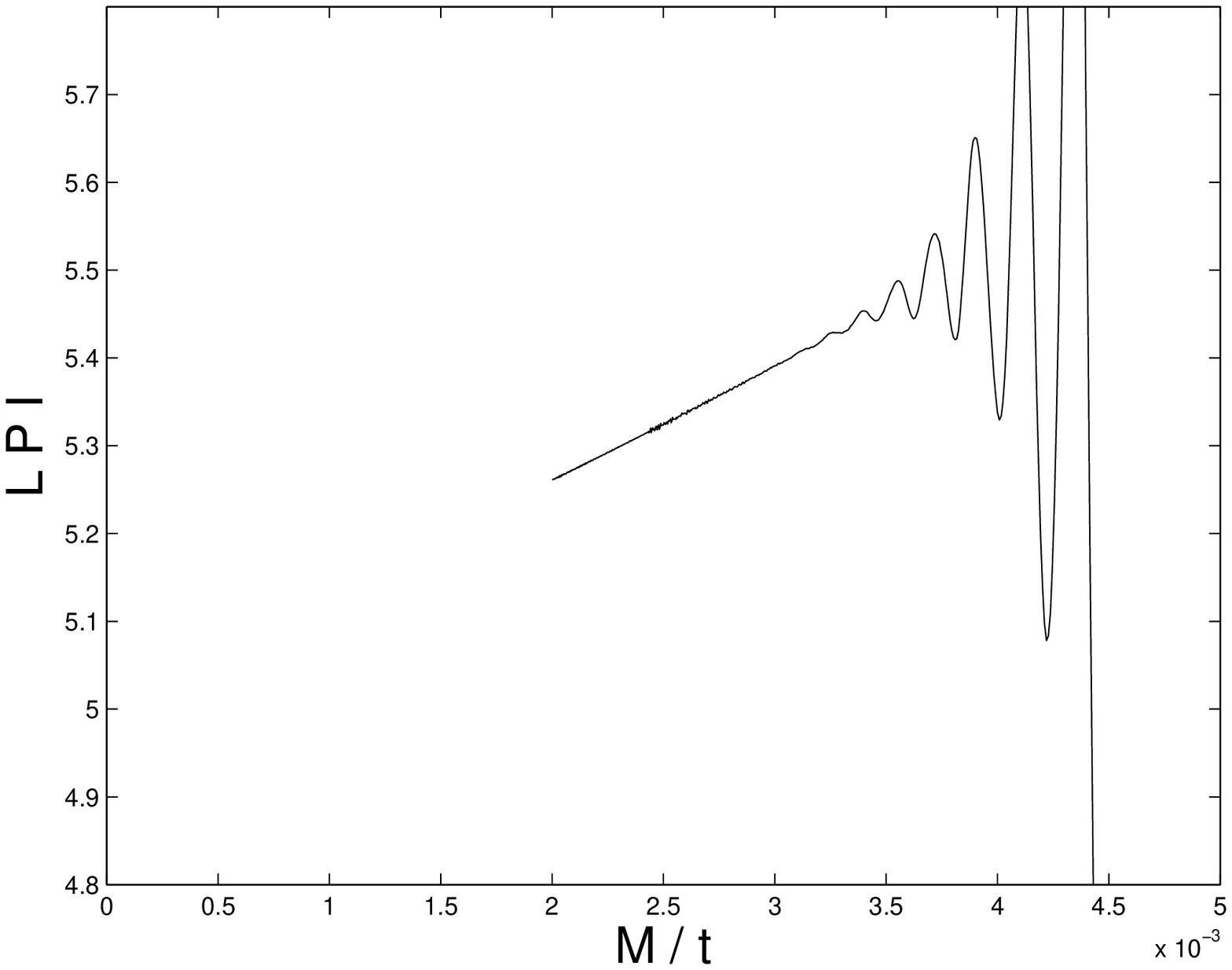} 
\caption{The local power index as a function of time for the monopole projection of the full field of Fig.~\ref{classB_4field}. When extrapolated to infinite time, the slope is $n_{\infty}=5.0129$.}
\label{classB_4lpi}
\end{figure}

Specifying initial data corresponding the a Class B hexadecapole ---namely, a pure $\ell=4$ moment on a Boyer--Lindquist time slice--- we find the field and it's multipole projections to behave as in Fig.~\ref{classB_4field}. Inspection of 
Fig.~\ref{classB_4field} shows we have not integrated long enough for the excited monopole to dominate the full field. We can nevertheless still find the asymptotic tail index. This is done by finding the local power index of the monopole projection of the field (shown in Fig.~\ref{classB_4lpi}), and extrapolating it to infinitely late times. We find the tail's slope to be $5.0129$, within $0.3\%$ from the value appropriate for Tail B.

\begin{figure}[htbp]
   \includegraphics[width=3.4in]{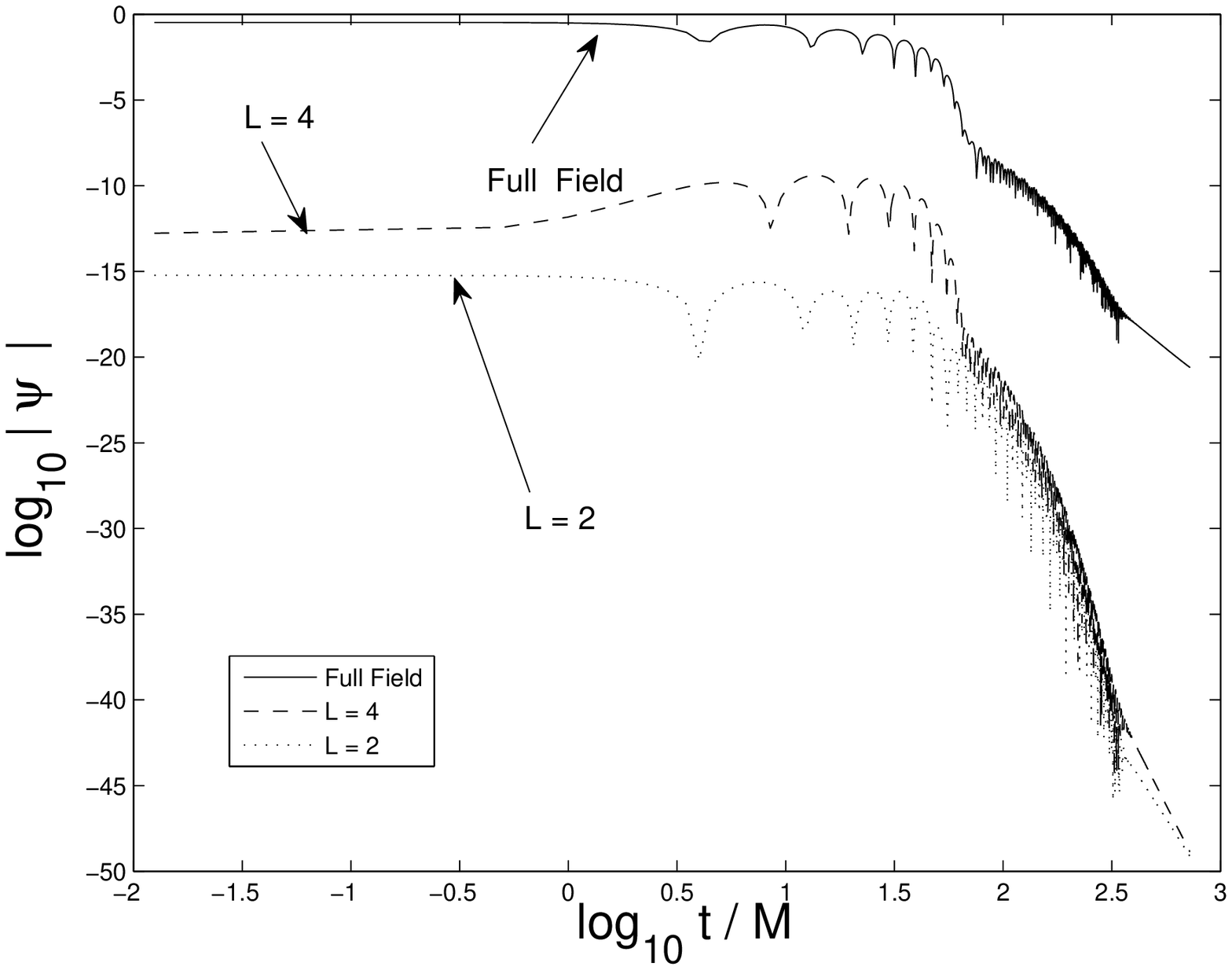} 
\caption{Evolution of Class B initial data for $\ell=6$. Grid parameters are: $\,\Delta\theta=1/48$, $\,\Delta r=M/32$. Shown are the full field (solid curve), and the hexadecapole (dashed curve) and quardupole (dotted curve) projections. (The monopole projection is not shown.) The initial data are again a gaussian centered at $r_*=25M$ with width of $4M$.}
\label{classB_6_0_2field}
 \includegraphics[width=3.4in]{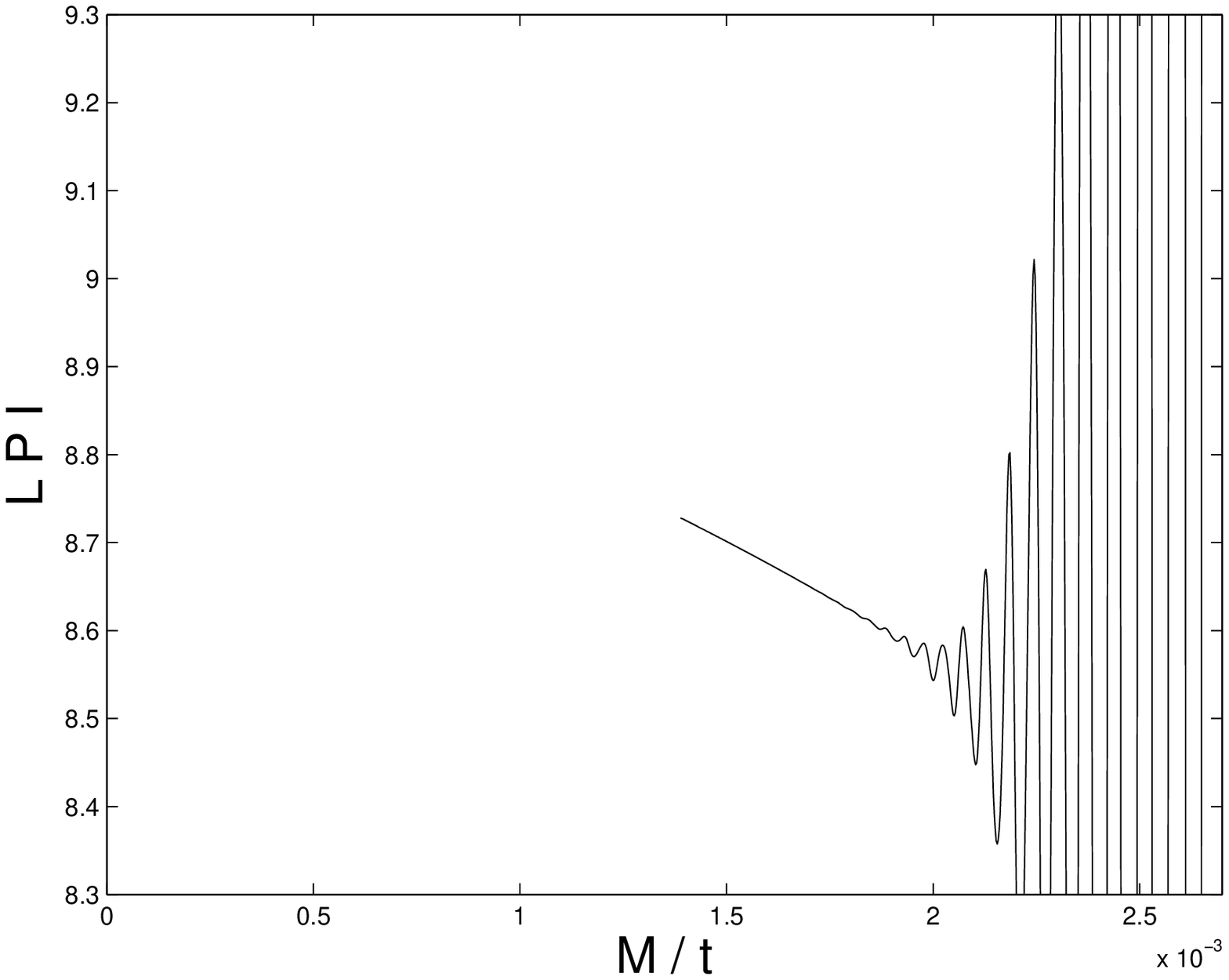} 
\caption{The local power index as a function of time for the quadrupole projection of the full field of Fig.~\ref{classB_6_0_2field}. When extrapolated to infinite time, the slope is $n_{\infty}=9.0568$.}
\label{classB_6_0_2lpi}
\end{figure}

Specifying the Class B initial data to $\ell=6$, we focus attention next on the subdominant quadrupole mode.  Figure \ref{classB_6_0_2field} depicts the full field and its hexadecapole and quadrupole moments, and Fig.~\ref{classB_6_0_2lpi} shows the local power index for the quadrupole projection. Extrapolating to infinite times, we find the tail's slope to equal 
9.0568, a deviation of $0.6\%$ from the decay rate for the tail would be, if the quadrupole were the least multipole to be excited. 

We therefore propose a generalization of the Tail B rule, that applies also for subdominant modes. Specifically, we propose 
that starting with an initial mode $\ell$, the decay rate of an (allowed) excited mode $\ell'$ is given by $n=\ell+\ell'+1$ if $\ell-\ell'>2$, and $n=2\ell'+3$ otherwise. This rule includes that of \cite{hod} as a particular case, when $\ell'=m,m+1$. As an example for the application of this generalized rule, consider the axisymmetric case above of  $\ell=6$ and $\ell'=2$ (which is a subdominant mode, as also the monopole is excited). The quadrupole mode $\ell'=2$ then indeed decays with $n=9$ as predicted by our proposal.

\subsection{Non--axisymmetric modes}

\begin{figure}[htbp]
   \includegraphics[width=3.4in]{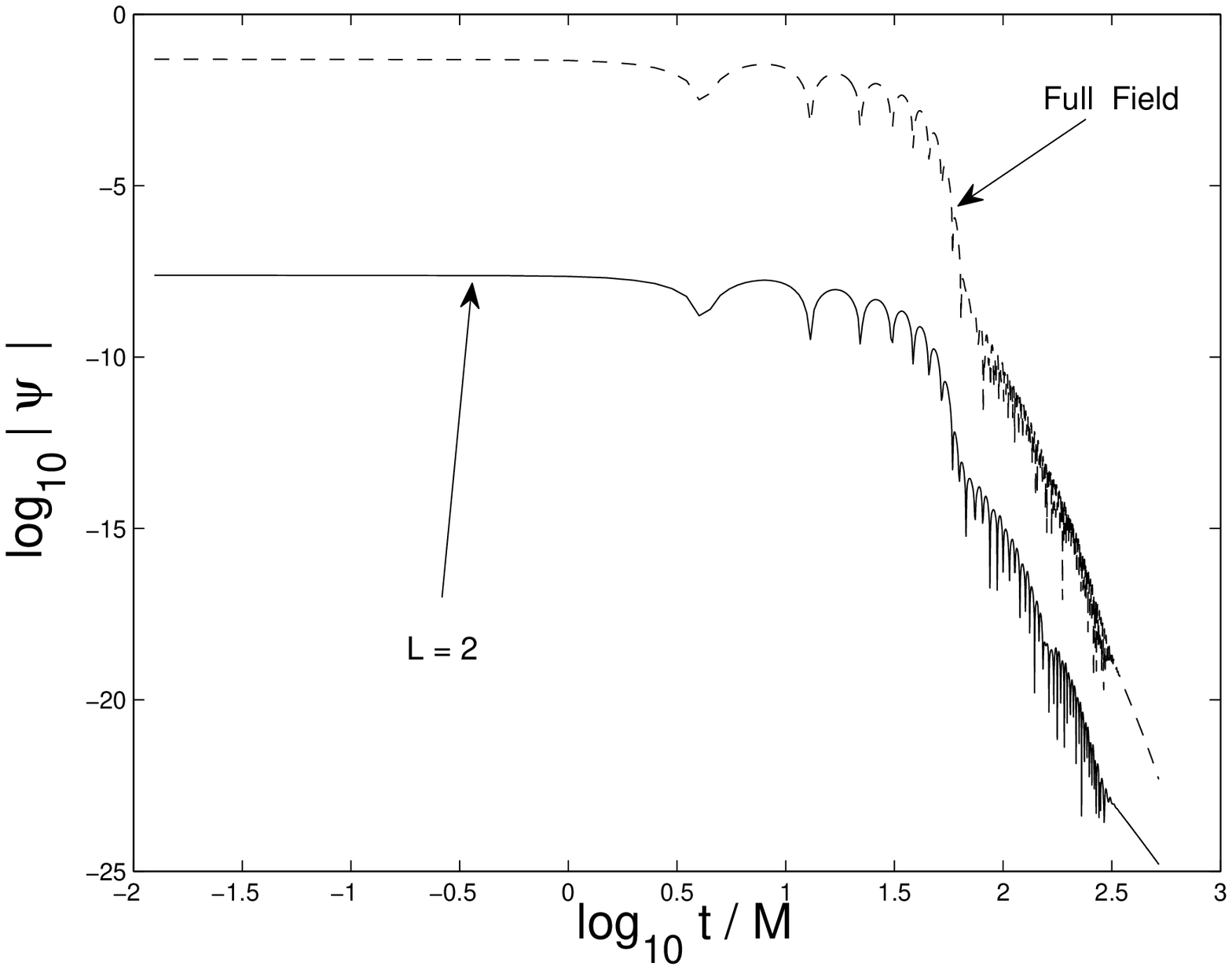} 
\caption{Evolution of Class B initial data for $\ell=6$ and $m=2$. Grid parameters are: $\,\Delta\theta=1/48$, $\,\Delta r=M/32$. Shown are the full field (solid curve), and the hexadecapole (dashed curve) and quardupole (dotted curve) projections. (The monopole projection is not shown.) The initial data are again a gaussian centered at $r_*=25M$ with width of $4M$.}
\label{classB_6_2field1}
 \includegraphics[width=3.4in]{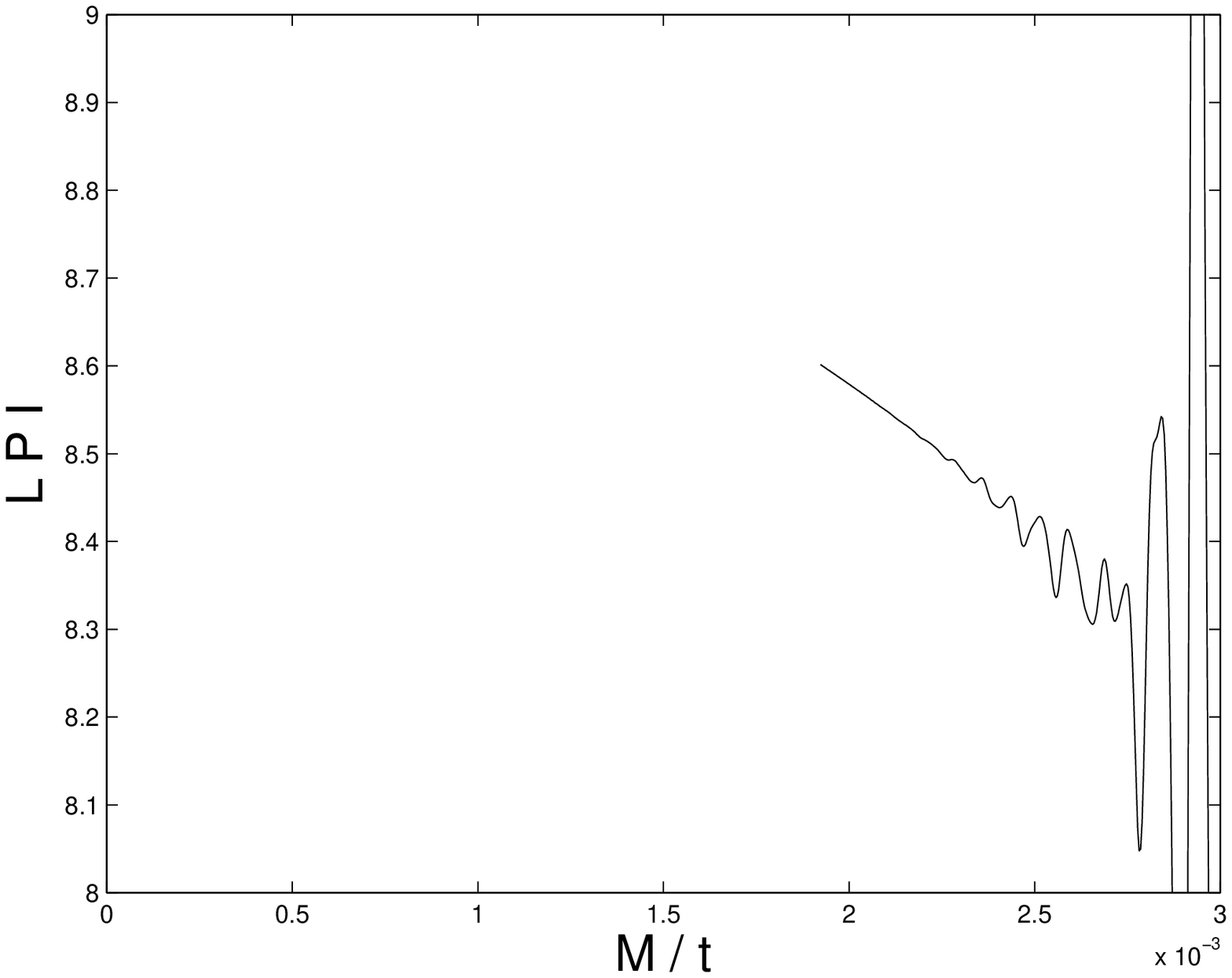} 
\caption{The local power index as a function of time for the quadrupole projection of the full field of Fig.~\ref{classB_6_2field1}. When extrapolated to infinite time, the slope is $n_{\infty}=8.91$.}
\label{classB_6_2lpi}
\end{figure}

We next consider the case of non-axisymmetric initial data of Class B. Specifically, consider initial $\ell=6$ and the azimuthal mode is $m=2$. In this case mode couplings allow the hexadecapole and quadrupole modes to be excited, but the monopole 
excitement is disallowed. According to our proposal, as $\ell'=2$, $\ell-\ell'=6-2=4>2$, and therefore $n=\ell+\ell'+1=6+2+1=9$. 
We show in Fig.~\ref{classB_6_2field1} the full field and the quadrupole mode as functions of time, and in Fig.~\ref{classB_6_2lpi} we plot the local power index for the quadrupole projection as a function of (inverse) time. We therefore conclude that our proposal for the power law indices predicts the correct value also for non-axisymmetric initial data. To our knowledge, this is the first time that non-axisymmetric Kerr tails are considered. 

\section{Optical illusion}\label{optical_illusion}

\begin{figure}[htbp]
   \includegraphics[width=3.4in]{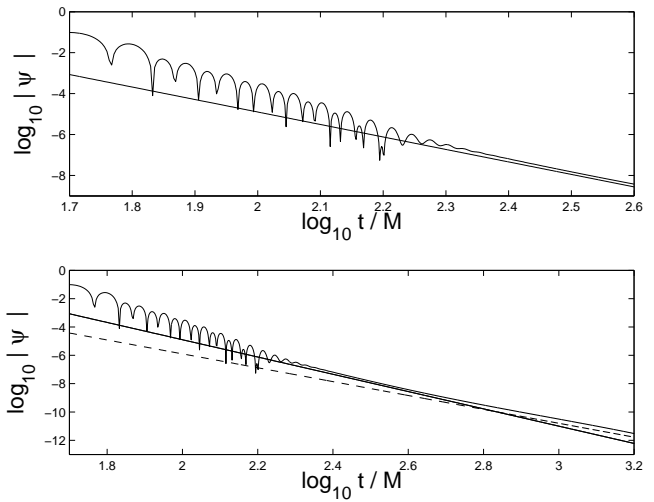} 
\caption{Solid curve: the full field. Solid reference line has a slope of 6.1. Dashed reference line has a slope of 4.9. Initial data were a Gaussian with $\ell=2$. It was centered at 30.0M with a width of 4M.
The grid resolution was: radial resolution $\,\Delta r=M/20$ and angular resolution $\,\Delta\cos\theta=1/32$. Here, $a/M=0.8$.
}
\label{OI1}
\end{figure}

When having asymptotic phenomena, it is crucial that the observations are made in the asymptotic regime. In Fig.~\ref{OI1} we show the field starting at the same times, but extending to two different values of time. In the upper panel extending to $t=400M$, and in the lower extending to $t=1600M$. In the upper panel we show a reference line with a slope of 6.1. One may be tempted to deduce from the seemingly two parallel curves at late times that the asymptotic slope of the field is 6.1. In the lower panel of Fig.~\ref{OI1} we show the same field and the same reference line, but we add another reference line, with slope of 4.9. Extending the integration time, it now appears that the asymptotic slope is 4.9. Which is it then? 6.1 or 4.9? The correct answer is 3, as we specified here Class A initial data. The slope of 5.5 reported on in \cite{krivan}, we believe, is just this effect: the field's decay rate was determined in \cite{krivan} not in its true asymptotic regime.

\begin{figure}[htbp]
   \includegraphics[width=3.4in]{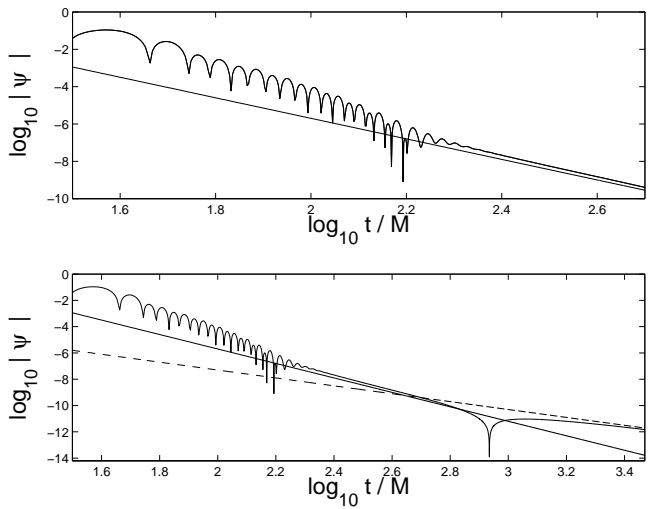} 
\caption{Solid curve: the full field.  Solid reference line has a slope of 5.5. Dashed reference line has slope of 3. Initial data were a Gaussian with $\ell=2$. It was centered at 17.5M with a width of 4M.
The grid resolution was: radial resolution $\,\Delta r=M/20$ and angular resolution $\,\Delta\cos\theta=1/32$.  Here, $a/M=0.8$.
}
\label{OI2}
\end{figure}

In Fig.~\ref{OI2} we show the full field up to $t=500M$ and to $t/M=2950$. At first we show a reference curve with slope of 5.5, and then increase the integration time, and show both the first reference curve, and a new one with slope of 3.0. In particular, if we were uncareful we could report on a tail index of 5.5 like that of \cite{krivan}. Continuing the simulation, however, we find the true asymptotic tail, with a tail index of 3, as expected. 

\begin{figure}[htbp]
   \includegraphics[width=3.4in]{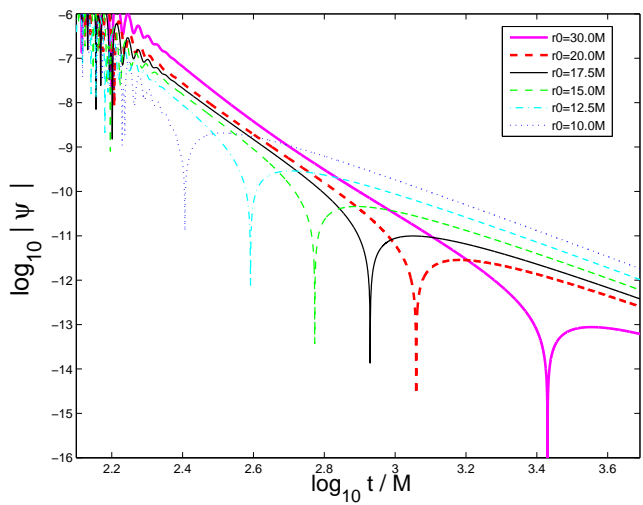} 
\caption{Effect of varying the location of the initial pulse. In all these runs, initial data was a Gaussian of width $\lambda=4M$, with $\ell=2$, and $a/M=0.8$. The grid resolution was: 0.05 (r) $\times$ 0.03125 (angle). The curves shown and the corresponding value for $r_0$, the center of the initial data gaussian, are: 
dotted curve: $r_0=10M$, dash-dotted curve: $r_0=12.5M$, dashed curve: $r_0=15M$, solid curve: $r_0=17.5M$, dashed thick curve: $r_0=20M$, and solid thick curve: $r_0=30M$. All data are extracted at $r=17.5M$. }
\label{OI3}
\end{figure}

The arbitrariness of the decay rate, as determined from intermediate time segments, is further exemplified in Fig.~\ref{OI3}, which shows the field for a number of different choices of $r_0$, the center of the initial pulse. The closer the pulse is to the black hole, the sooner the true asymptotic tail appears. Conversely, pushing the initial pulse outwards, we protract the intermediate tail regime, and can make it as long as we want: it is fully controllable. The choice of the length of the intermediate tail also affects its decay rate determination. We therefore see that not only is such a decay rate meaningless, it is also directly manipulable. The important lesson here, reflecting on the claims made in \cite{krivan}, is that a distant initial pulse makes it harder to see the asymptotic tail, and it would require longer time evolutions.  In \cite{krivan}, the initial pulse was chosen at $r_0=100M$, which would make the intermediate tail very protracted indeed.

\begin{figure}[htbp]
   \includegraphics[width=3.4in]{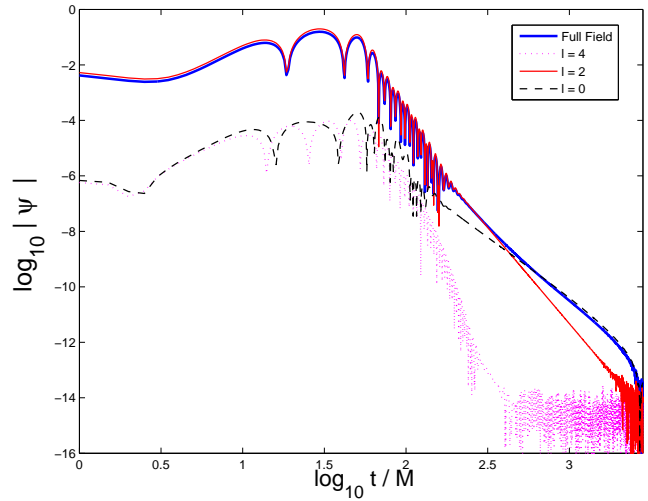} 
\caption{Same data as in Fig.~\ref{OI1}. The full field is decomposed into the Boyer--Lindquist multipole modes.  }
\label{OI4}
\end{figure}

Figure \ref{OI4} explains the broken power-law behavior we see in Fig.~\ref{OI1}, and also why neither is the true asymptotic field: Up to $t\sim 500M$ the field is dominated by the dipole ($\ell=2$) mode, and at later times it is dominated by the monopole ($\ell=0$). Notice, that even the seemingly constant slope in the lower panel of Fig.~\ref{OI1} does not represent the asymptotic behavior of the field, as is clear from Fig.~\ref{OI4}. 
The ``optical illusion" effect is not a result of mode mixing: it appears in each mode separately. Indeed, in the $\ell=0$ mode what may appear as a straight segment (on the log--log plot of Fig.~\ref{OI4}) is curving if the integration time is extended.  But when we combine the two separate optical illusions of the $\ell=2$ and $\ell=0$ modes, we may get a broken power--law as in the lower panel of Fig.~\ref{OI1}, that has nothing to do with the true asymptotic decay rate of the field. Notice, that also the $\ell=4$ is excited, but at a negligible level. Higher (even) modes are also excited, but not shown in Fig.~\ref{OI4}.

\begin{figure}[htbp]
   \includegraphics[width=3.4in]{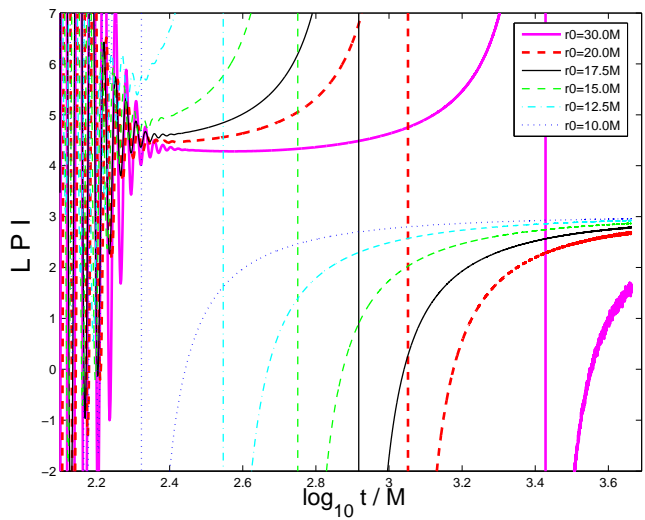} 
\caption{Local power index (LPI) as a function of $t$ for the same data as in Fig.~\ref{OI3} for the $\ell=0$ Boyer--Lindquist multipole mode.}
\label{OI5}
\end{figure}

\begin{figure}[htbp]
   \includegraphics[width=3.4in]{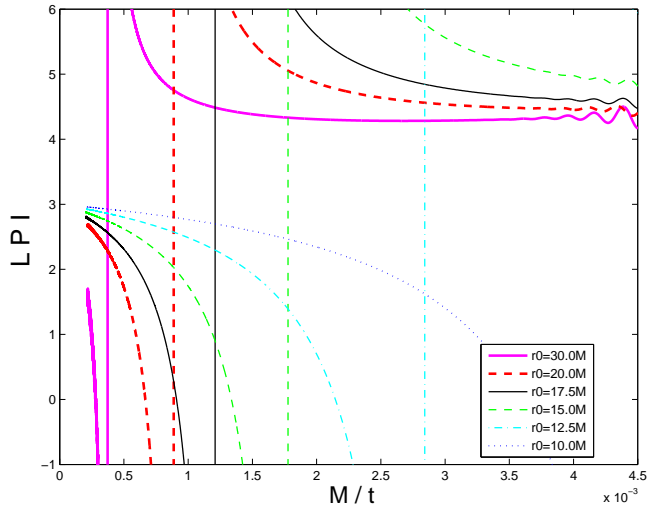} 
\caption{Same as Fig.~\ref{OI5}, but presented as a function of $M/t$. }
\label{OI6}
\end{figure}

In view of he arbitrariness in the slope as calculated over some finite range, it appears that a better way to evaluate the slope is through its local value, the {\em local power index} defined by $n(t):=-{\dot \psi}\,t/\psi$ \cite{burko-ori}. Even plotting the local power index as a function of the time could be misleading, as is evident from Figs.~\ref{OI5} and \ref{OI6}: notice how the bold solid curve appears to approach a constant value for low values of $t$. Indeed, changing the initial conditions, one can made the local power index have an arbitrarily protracted intermediate value. 

\begin{figure}[htbp]
   \includegraphics[width=3.4in]{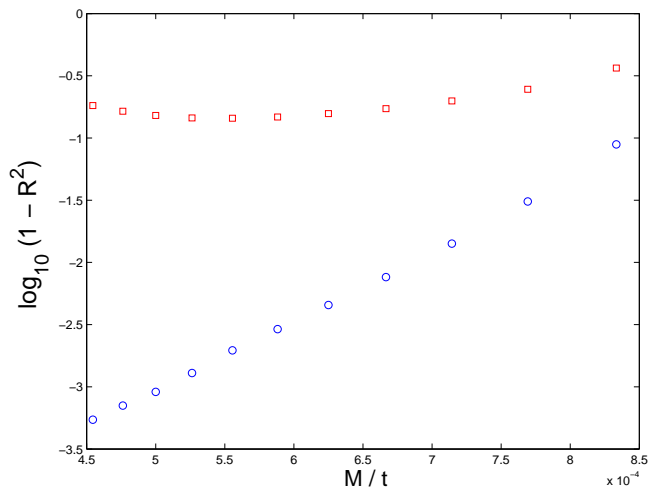} 
\caption{Dependence of the squared correlation coefficient $R^2$ on the {\em local} temporal location of an interval of constant lapse (lasting $1000M$). Data are taken for the same initial data as in Fig.~ \ref{OI3}. For both cases the data correspond to the squared correlation coefficient of ten evenly spaced (in $t/M$) values of the local power index (with increments of $100M$) ending at the value shown, fitted to the ansatz (\ref{ansatz}): Squares ($\square$) correspond to data for $r_0=30M$, and circles ($\circ$) to $r_0=10M$. }
\label{OI7}
\end{figure}

How can then one distinguish an arbitrarily protracted intermediate behavior form the true asymptotic behavior? The key point in distinguishing the true asymptotic tail from an intermediate one is to consider the asymptotic behavior of the local power index. As shown in \cite{smith-burko}, the asymptotic behavior of the local power index is 
\begin{equation}\label{ansatz}
n(t)=n_{\infty}+n_1\,(M/t)+\cdots\, .
\end{equation} 
One should therefore test how well the fit of $n(t)$ as a function of $M/t$ improves with $t$. Loosely speaking, intermediate regions, like those seen in Fig.~\ref{OI6}, will generally yield fits that deteriorate with time, until the true asymptotic regime is reached. How should this criterion be applied in practice? A {\em necessary condition} is that the ansatz  (\ref{ansatz}) is satisfied {\em locally}. Specifically, Fig.~\ref{OI7}, displays the squared correlation coefficient $1-R^2$ for constant duration intervals (of ten equally spaced in time values of the local power index). In practice, we choose two sets of data differing only in the location of the initial data pulse, so that one has a protracted intermediate tail, and the other arrives quickly at the asymptotic regime. Indeed, not only does the squared correlation coefficient indicate only relatively poor agreement with the ansatz (\ref{ansatz}) for the intermediate tail, it also fails to improve at later times. In contrast, at the true asymptotic regime, the squared correlation coefficient approaches unity consistently. We are therefore motivated to introduce the following definition:

\noindent 
{\bf Definition}: The local power index satisfies the ansatz (\ref{ansatz}) {\em locally} if the squared correlation coefficient for fitting the local power indices of constant duration intervals to the ansatz (\ref{ansatz}) approaches unity as $t$ increases.

The local satisfaction of the ansatz (\ref{ansatz}) is only a necessary condition for the true asymptotic tail, because in practice we can only examine numerically finite time values. Indeed, this is the very reason why misidentifying an intermediate tail as an asymptotic one is a problem in the first place. Perhaps not surprisingly, one can find intermediate tails for which the ansatz  (\ref{ansatz}) appears to be satisfied locally, but is violated {\em globally} in the following sense. 

\noindent 
{\bf Definition}:  The local power index satisfies the ansatz (\ref{ansatz}) {\em globally} if the squared correlation coefficient for fitting the local power indices of consecutively longer intervals to the ansatz (\ref{ansatz}) starting from an arbitrary (late) time approaches unity as $t$ increases.

\begin{figure}[htbp]
   \includegraphics[width=3.4in]{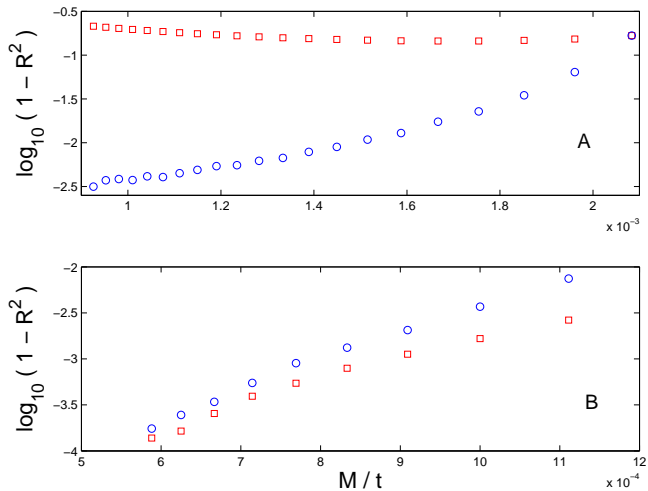} 
\caption{Dependence of the squared correlation coefficient $R^2$ on a {\em global} temporal interval. 
Upper panel (A): behavior of an intermediate tail. Lower panel (B): behavior of an asymptotic tail. 
In both panels the local satisfaction of ansatz (\ref{ansatz}) is shown with circles ($\circ$), and the global violation (A) or satisfaction (B) of the same ansatz is shown with squares ($\square$). Initial data are centered on $r_0=30M$ (same as in Fig.~\ref{OI3}). Data for the upper panel (A) correspond to five evenly spaced (in $t/M$) values of the local power index (with increments of $30M$) ending at the value shown (local) and appropriately many values with the same spacing (starting at $t=360M$) and ending at the shown time. Data for the lower panel (B) correspond to five evenly spaced (in $t/M$) values of the local power index (with increments of $100M$) ending at the value shown (local) and appropriately many values with the same spacing (starting at $t=500M$) and ending at the shown time.  }
\label{OI8}
\end{figure}

\noindent In Fig.~\ref{OI8} we show the local satisfaction but global violation of ansatz (\ref{ansatz}) for an intermediate tail (upper panel (\ref{OI8}A)), and their simultaneous satisfaction for the asymptotic tail (lower panel (\ref{OI8}B)). 

We are therefore motivated to propose the following as a criterion for distinguishing intermediate from asymptotic tails:

\noindent 
{\bf A criterion for identification of the asymptotic tail}:  The tail is asymptotic if the ansatz (\ref{ansatz}) is satisfied simultaneously both locally and globally.

While figures such as Fig.~\ref{OI1} or the upper panel of Fig.~\ref{OI2} may lead to a misidentification of an intermediate tail as an asymptotic one, we believe that our criterion will expose an intermediate tail as such, and correctly identify asymptotic tails.

\section{Ingoing Kerr slicing}\label{s_ingoing_kerr}

The tails for initial data specified on ingoing Kerr slices were first discussed in \cite{burko-khanna}. The initial data used in \cite{burko-khanna} are of Class A, leading to Tail A behavior. Recently, it has argued that a pure multipole in the ingoing Kerr coordinates does not correspond to a pure multipole in Boyer--Lindquist coordinates, and that in addition the different slicing conditions in the two cases were the reason that \cite{burko-khanna} found Tail A behavior and not Tail B behavior. 

Ingoing Kerr coordinates are different from Boyer--Lindquist coordinates by a transformation of the $\phi$ and $t$ coordinates ---with the transformation depending on the radial coordinate only--- leaving the ingoing Kerr $r,\theta$ coordinates identical to their Boyer--Lindquist counterparts. Specifically, the transformation is 
${\breve \phi}={\hat\phi} + a \int {\hat \Delta}^{-1}\,d{\hat r}$ and ${\breve t}={\hat t}+{\hat r}_*-{\hat r}$, where ${\breve t},{\breve\phi}$ are the ingoing Kerr coordinates, ${\hat t},{\hat\phi}$ are the Boyer--Lindquist coordinates, and where ${\hat\Delta}={\hat r}^2-2M{\hat r}+a^2$ and ${\hat r}_*=\int({\hat r}^2+a^2){\hat \Delta}^{-1}\,d{\hat r}$. The different sets of coordinates we use are summarized in Table \ref{table1}.

\begin{table}[htdp]
\caption{The notation used for the different coordinate systems and the transformations between them. See the text for more detail.}
\begin{center}
\begin{tabular}{||l|c|c||}
\hline\hline
Coordinates  & Notation & Transformation \\
\hline\hline
Boyer--Lindquist & ${\hat t},{\hat r},{\hat \theta},{\hat \phi}$ & \\
\hline
Ingoing Kerr & ${\breve t},{\breve r},{\breve \theta},{\breve \phi}$ &  ${\breve \phi}={\hat\phi} + a \int {\hat \Delta}^{-1}\,d{\hat r}$ \\
 & & ${\breve t}={\hat t}+{\hat r}_*-{\hat r}$ \\
\hline
Poisson & ${\tilde t},{\tilde r},{\tilde \theta},{\tilde \phi}$ & ${\tilde r}=\sqrt{{\hat r}^2+e^2\sin^2{\hat\theta}}$ \\
 & & $\cos^2{\tilde\theta}=\frac{{\hat r}^2\cos^2{\hat\theta}}{{\hat r}^2+e^2\sin^2{\hat\theta}}$ \\
\hline
\end{tabular}
\end{center}
\label{table1}
\end{table}%

A spherical harmonic mode in Boyer--Lindquist coordinates with a specific multipole  $\ell$ and a specific azimuthal mode $m$ is therefore also a spherical harmonic mode in ingoing Kerr coordinates with the same values for $\ell,m$, and the difference is only in a different function of the radial coordinate. One may therefore expect that initial data specified in ingoing Kerr coordinates would lead to the same tail behavior as in Boyer--Lindquist coordinates. One may object to this expectation by  stating that the constant ${\hat t}$ slices in the Boyer--Lindquist case are not the same as the constant ${\breve t}$ slices in the ingoing Kerr case. While this difference is certainly existing and the two hypersurfaces are not the same, we find that it does not change the tail behavior. 

\begin{figure}[htbp]
   \includegraphics[width=3.4in]{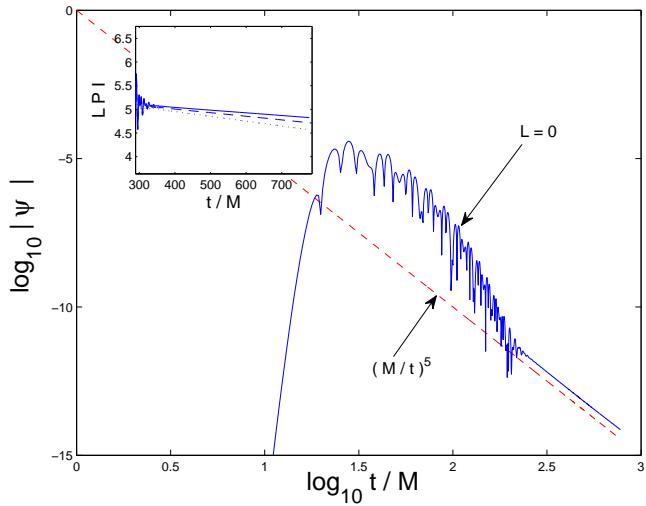} 
\caption{Evolution of Class B initial data of a pure azimuthal scalar field hexadecapole ($\ell=4$) mode in ingoing Kerr coordinates on a ${\breve t}={\rm const}$ slice. Here, $a/M=0.8$. The initial data radial profile is a gaussian with ${\breve r}_0=2M$ and ${\breve \lambda}=1M$. The grid resolution is $\,\Delta{\breve\theta}=1/64$ and $\,\Delta {\breve r}=2M/125$. The dashed curve is the carve $\psi=(M/{\breve t})^5$ which is added for reference. The local power index is calculated for three radial resolutions: $\,\Delta {\breve r}=M/40$ (dotted), $\,\Delta {\breve r}=M/50$ (dashed), and $\,\Delta {\breve r}=2M/125$ (solid).}
\label{ingoing_kerr}
\end{figure}

Class A initial data were studied in detail in \cite{burko-khanna}. We therefore present here the results for Class B initial data. We find that starting with a pure $\ell=4$ azimuthal mode (in ingoing Kerr coordinates on a ${\breve t}={\rm const}$ slice) the tail has Tail B behavior, similar to that of Boyer--Lindquist coordinates and Boyer--Lindquist slicing. Figure \ref{ingoing_kerr} shows the field and the local power index for these data. We infer from the simulation data that the tail index is $5.088$, a deviation of $1.7\%$ from the Tail B prediction of 5. 

The ingoing Kerr results are obtained using a code based on the Penetrating Teukolsky Code (PTC) \cite{ptc}. The PTC is written on a fixed ingoing Kerr coordinate grid, which means that grid resolution becomes increasingly poor approaching the event horizon. While simple Fixed Mesh Refinement methods can resolve this difficulty, we were able to obtain the tail index using the following method: we first find the local power index for a number of time values for different choices of the radial (and temporal) resolution. We then extrapolate the local power indices to infinite radial resolution, and then extrapolate the sequence of local power indices at different values of the time (and with infinite resolutions) to infinitely late times. The result is rather close to the expected value of 5, but suffers from somewhat higher numerical error because of the two extrapolation cycles. As noted above, the deviation of the tail index  from the Tail B value is at the order of $10^{-2}$. We therefore conclude that Class B initial data in the ingoing Kerr case lead to Tail B behavior.

We therefore contend that initial time slices, even when are distinct, fall into equivalency classes in terms of the type of tail behavior that emerges (recall that on a given slice multipoles can also mix because of coordinate transformations). The Boyer--Lindquist and ingoing Kerr slices belong to the same equivalency class, in the sense that one finds the same Tail B behavior for Class B initial data. The Kerr--Schild slices belong to a distinct equivalency class, and Class B initial data lead to Tail A behavior. We conjecture that the Boyer--Lindquist and ingoing Kerr slices belong to the same equivalency class because the time slices in either are related by a transformation that does not involve the polar angular coordinate and that depends only on the temporal and azimuthal angular coordinate. One may therefore conjecture that other slicing conditions that differ from the Boyer--Lindquist slicing condition by similar transformations will behave accordingly.

\section{poisson data}\label{poisson_data}

Poisson has made the argument that if Class B initial data are specified in so-called ``spherical coordinates," rather than in Boyer--Lindquist coordinates, which Poisson calls ``spheroidal," mode evolution is changed \cite{poisson}. In particular, Poisson proposes that in Kerr spacetime such initial data lead to Tail B behavior. In a spacetime with global weak curvature (no black hole, but no assumptions on symmetry), Poisson data indeed lead to such late--time behavior. This may be surprising, as a pure mode of Poisson coordinates may be written as a mixed state of Boyer--Lindquist modes. This mixed state of modes effectively behaves as Class A initial data, so that the intuitively expected behavior is that of Tail A. 

Clearly, the assumption of globally weak curvature affects the tails, and shifts them from their Tail A values to those reported by Poisson. This could be the case if the field reflects off the regular center in the Poisson spacetime,  so that the leading order part of the field (in $t^{-1}$) is cancelled out and one is left with the next order in $t^{-1}$, or if the monopole part of the field is canceled after the reflection, leaving only the dipole and higher moments present \cite{GPP}.  The following question arises: what are the tail indices for ``pure Poisson" Class B data in a Kerr spacetime? Specifically, if one sets at $t=0$ a pure Class B hexadecapole mode with respect to the Poisson ``spherical" coordinates, does the late--time tail decay as $1/t^3$, as appropriate for Tail A,  or as $1/t^5$ as appropriate for Tail B (and as predicted by \cite{poisson})? As we are showing here, the answer is clearly $1/t^3$. 

\begin{figure}[htbp]
   \includegraphics[width=3.4in]{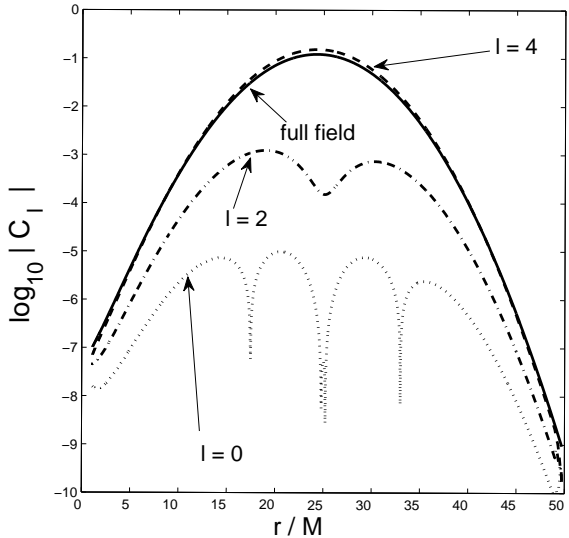} 
\caption{A scalar field ``pure Poisson" $\ell=4$ mode, ${\tilde C}_4({\tilde r})$ (solid curve), and its projections on Boyer--Lindquist $\ell$ modes $_4{\hat C}_{\ell'}({\hat r})$: dashed ($\ell'=4$), dash-dotted ($\ell'=2$), and dotted ($\ell'=0$).  Here, $a/M=0.995$. The initial data radial profile is a gaussian with ${\tilde r}_0=25M$ and ${\tilde \lambda}=4M$. The abscissa is ${\tilde r}/M$ or ${\hat r}/M$ as is appropriate.}
\label{PD1}
\end{figure}

Initial data that describe a ``pure Poisson" mode are a linear combination of Boyer--Lindquist modes, as one may expand such a mode in the function space of Boyer--Lindquist spherical harmonics. Importantly, as the transformation from Boyer--Lindquist to Poisson coordinates does not involve the time, Boyer--Lindquist slices are identical to Poisson slices. 
Therefore, one may use a code in which the wave equation is written in Boyer--Lindquist coordinates to simulate the evolution of a ``pure Poisson" mode, if one writes the initial data as the appropriate linear combination of Boyer--Lindquist modes. This may be done as follows:
Denote the Boyer--Lindquist coordinates as ${\hat r},{\hat \theta}$ and the Poisson ``spherical" coordinates as ${\tilde r},{\tilde\theta}$. As shown in \cite{poisson}, the coordinate transformation is 
\begin{eqnarray}
{\tilde r}&=&\sqrt{{\hat r}^2+e^2\sin^2{\hat\theta}}\\
\cos^2{\tilde\theta}&=&\frac{{\hat r}^2\cos^2{\hat\theta}}{{\hat r}^2+e^2\sin^2{\hat\theta}}\, ,
\end{eqnarray}
where $e$ is a measure of the ``ellipticity" of the Boyer--Lindquist coordinates, which in practice equals the Kerr black hole's spin $a$. To specify a ``pure Poisson" mode, we take the field to be 
\begin{eqnarray}
\left.\psi_{\ell}\right|_{t=0}&=&{\tilde C}_{\ell}({\tilde r})\,P_{\ell}(\cos{\tilde\theta})\nonumber \\
&\equiv& A\,\exp\left(-\frac{({\tilde r}-{\tilde r}_0)^2}{2{\tilde\lambda}^2}\right)\,P_{\ell}(\cos{\tilde\theta})\, ,
\end{eqnarray}
after choosing the radial function to be a gaussian of width ${\tilde\lambda}$ centered on ${\tilde r}_0$. 
The initial ``pure Poisson" field $\left.\psi_{\ell}\right|_{t=0}$ can be expanded in the function space of  Boyer--Lindquist Legendre polynomials as 
\begin{equation}
\left.\psi_{\ell}\right|_{t=0}=\sum_{\ell'=0}^{\infty}\,_{\ell}{\hat C}_{\ell'}({\hat r})\,P_{\ell'}(\cos{\hat\theta})\, .
\end{equation}
As shown in \cite{poisson}, the functions $_{\ell}{\hat C}_{\ell'}({\hat r})$ are complicated. (In fact, only the coefficients for the {\em inverse} transformation are found in \cite{poisson} (for $\ell=4$ and $\ell'=0$) as a power series in the ``ellipticity" $e$.) In Fig.~\ref{PD1} we show the functions ${\tilde C}_4({\tilde r})$, and $_4{\hat C}_{\ell'}({\hat r})$ for $\ell'=0,2,4$ (in general, these functions are non-vanishing for all values of $\ell'$, respecting the parity symmetry), for a high value for the ``ellipticity," specifically for $e=0.995M$. To complete the initial data problem, we also need to specify at $t=0$ the time derivative of the field. 
For Class B initial data we choose $\,\partial_t\psi=\,\partial_{{\hat r}_*}\psi$, and for Class A initial data we choose $\Pi=0$.

\begin{figure}[htbp]
   \includegraphics[width=3.4in]{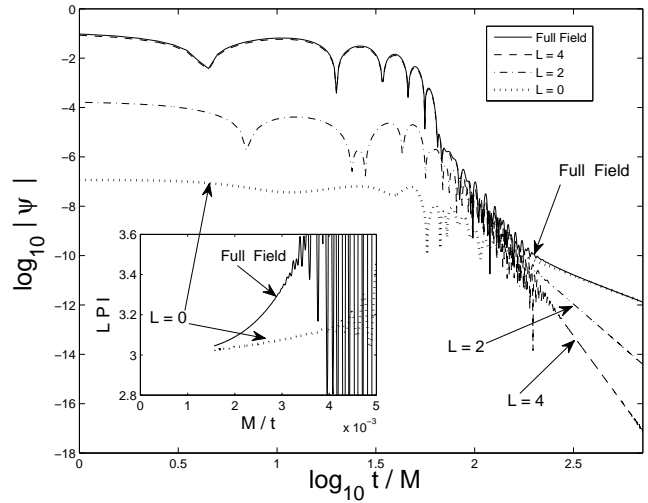} 
\caption{Time evolution of a Class B ``pure Poisson" $\ell=4$ mode, and its Boyer--Lindquist $\ell'=0,2,4$ modes, and the local power indices for the full field and for the Boyer--Lindquist monopole projection. Here, $a/M=0.995$. The (Class B) initial data are those of Fig.~\ref{PD1}.}
\label{Poisson_ClassB}
\end{figure}

Figure \ref{Poisson_ClassB} depicts the full field and its multipole projections and the local power indices for the full field and its monopole projection as functions of time for Class B initial data. The late time behavior is clearly that of Tail A, with the monopole's decay rate at infinite time extrapolated to $3.0017$, within $6\times 10^{-4}$ from the predicted value of $3$.

\begin{figure}[htbp]
   \includegraphics[width=3.4in]{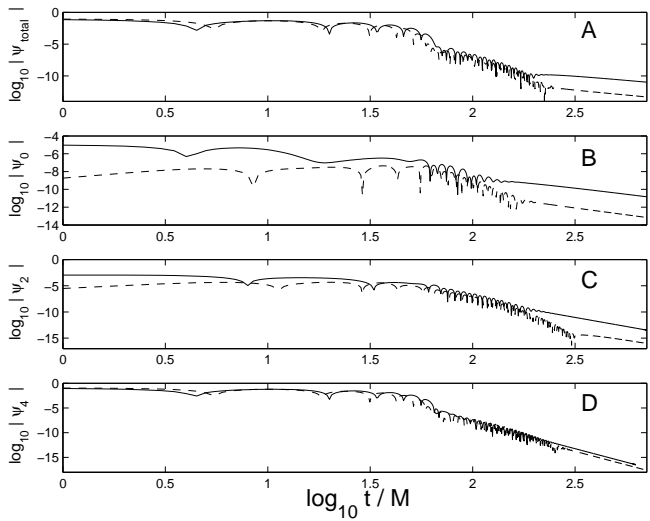} 
\caption{Time evolution of a Class A ``pure Poisson" $\ell=4$ mode (panel A), and its Boyer--Lindquist $\ell'=0,2,4$ modes (panels B,C, and D, respectively) (solid curves), and the evolution of the Boyer--Lindquist $\ell'=4$ projection of the ``pure Poisson" mode and its Boyer--Lindquist projections (dashed curves). Here, $a/M=0.995$. The initial data are those of Fig.~\ref{PD1}.}
\label{PD2}
\end{figure}

In Fig.~\ref{PD2} we consider Class A initial data. Again, it suggests that the tail field drops of as $1/t^3$ for a scalar field with initial data that are a ``pure Poisson" hexadecapole mode. Specifically, we show this to be the case more precisely, using the criterion based on simultaneous local and global behavior of the local power index. Comparison of the evolution of the evolution of the ``pure Poisson" $\ell=4$ mode with the evolution of its Boyer--Lindquist $\ell'=4$ projection suggests that the monopole projection $\ell'=0$ at the initial time is important at late times. The $\ell'=0$ mode is being excited during the evolution also when it is not present at the initial time, and eventually it dominates the field at late times. However, the generated $\ell'=0$ mode starts dominating the total field only at a later time, as its amplitude is lower than that of the monopole part of the full $\ell=4$ mode. The latter mode includes a monopole component already at the initial time, and this part of the field starts dominating the total field at an earlier time than the monopole component that is excited. For this reason the asymptotic  $\ell=4$ mode is much stronger than the asymptotic tail of initial data that include only the $\ell'=4$ projection of the $\ell=4$ mode, even though at the initial time they are almost identical, as can be seen in Fig.~\ref{PD1}. 

\begin{figure}[htbp]
   \includegraphics[width=3.4in]{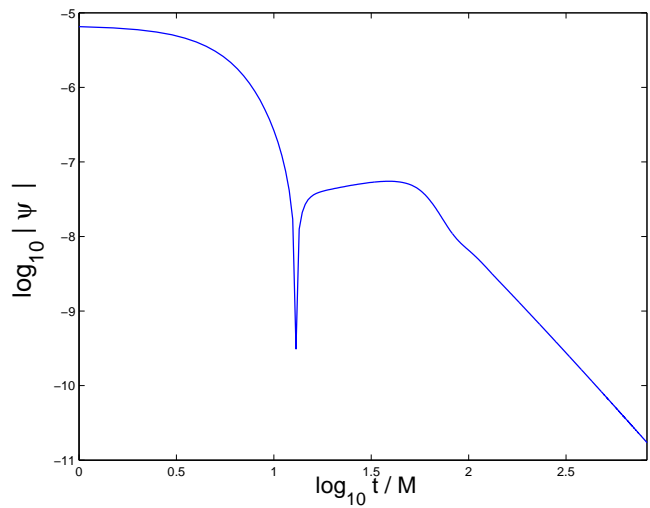} 
\caption{Time evolution of the Boyer--Lindquist $\ell'=0$ projection of the Class A ``pure Poisson" $\ell=4$ mode. The initial data are those of Fig.~\ref{PD1}.}
\label{PD3}
\end{figure}

To show that indeed the late--time tail is dominated by the evolution of the $\ell'=0$ projection of the original ``pure Poisson" $\ell=4$ mode at $t=0$, we present in Fig.~\ref{PD3} the evolution of this mode. Comparison with Fig.~\ref{PD2} indeed shows that the late time field of the full $\ell=4$ mode is dominated by the evolution of its monopole projection, not the monopole mode that is excited during the evolution because of mode coupling.

\begin{figure}[htbp]
   \includegraphics[width=3.4in]{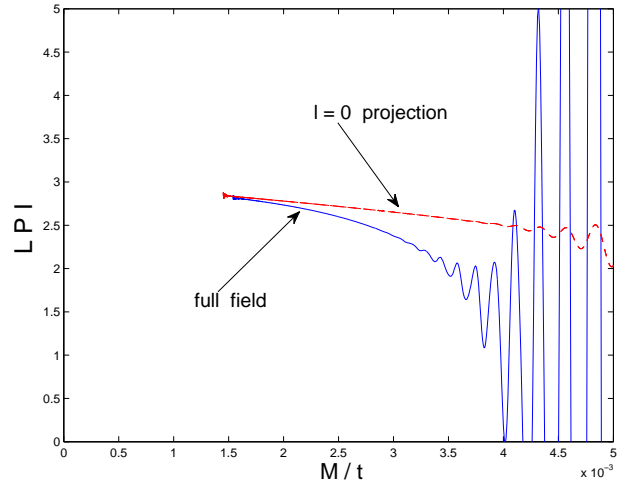} 
\caption{The local power index as a function of $t^{-1}$ for the Class A ``pure Poisson" mode $\ell=4$ (solid curve) and for its $\ell'=0$ Boyer--Lindquist projection (dashed curve). The initial data are those of Fig.~\ref{PD1}, and the field evolution is presented in Fig.~\ref{PD2}.}
\label{PD4}
\end{figure}


In Fig.~\ref{PD4} we show the behavior of the local power indices for the full field of the Class A ``pure Poisson" $\ell=4$ mode, and for its monopole $\ell'=0$ projection. Figure \ref{PD4} strongly suggests that the full field, and also the monopole projection, drop off at late times as $1/t^3$. 
Indeed, the tail seen in Figs.~\ref{PD3} and \ref{PD4} is asymptotic, and the local power index satisfies the ansatz  (\ref{ansatz}) both locally and globally simultaneously. 

As noted in \cite{poisson}, it is not surprising that starting with the initial data of a $Y_{\ell=4}^{m=0}$ mode, the 
$Y_{\ell=0}^{m=0}$ mode is excited. Figure \ref{PD2} (in conjunction with the conclusion that the tail is asymptotic) shows that when starting with a ``pure Poisson" mode, the excited monopole mode is insignificant for the asymptotic tail problem, as it is swamped by the evolution of the monopole projection of the initial data. Above, we showed that starting with Class B initial data of a pure Boyer--Lindquist $\ell=4,m=0$, the late time tail drops off according to $t^{-5}$, in accordance with Tail B behavior, consistently with the results of \cite{GPP,TKT}. In that case, there is no monopole field present at $t=0$, and all the monopole field present at late times is due to the excitation of the $\ell=0$ mode due to mode coupling.  The excited mode that has Tail B behavior is overwhelmed by the evolution of the monopole mode existing at the initial time that has Tail A behavior. 
Most importantly, we choose the initial data to be a pure spherical harmonic mode in the so-called ``spherical" coordinates. The late time tail still decays according to Tail A behavior, in contrast with the Tail B expectations based on \cite{poisson}. In the case presented, as the initial mode is azimuthal, the asymptotic tail drops off as $t^{-3}$.

\section*{acknowledgments}
The authors wish to thank Richard Price and Jorge Pullin for invaluable discussions. Early work on this problem was done with Elspeth Allen at the University of Utah. LMB is supported in part by  NASA/GSFC grant No.~NCC--580 and by NASA/SSC grant No.~NNX07AL52A. GK is grateful for research support from the UMD College of Engineering RSI Fund and SCEA. Many of the simulations were performed on NSF's TeraGrid infrastructure (under grant number TG-PHY060047T) and the HPC Consortium's facilities.

\end{document}